\documentclass[numberedappendix,iop]{emulateapj}
\def\bmath#1{\mbox{\boldmath$#1$}}
\def\revise#1{#1}

\slugcomment{Accepted for publication in ApJ}

\shorttitle{Collapse of magneto-turbulent cloud cores} 
\shortauthors{Matsumoto and Hanawa}

\begin{document}

\title{Protostellar collapse of magneto-turbulent cloud cores: shape during
  collapse and outflow formation}

\author{Tomoaki Matsumoto\altaffilmark{1} and Tomoyuki Hanawa\altaffilmark{2}}

\altaffiltext{1}{Faculty of Humanity and Environment, Hosei
 University, Fujimi, Chiyoda-ku, Tokyo 102-8160, Japan}
\email{matsu@hosei.ac.jp}

\altaffiltext{2}{Center for Frontier Science, Chiba University, 1-33, Yayoi-cho, Inage-ku, Chiba 263-8522, Japan}

\begin{abstract}
  We investigate protostellar collapse of molecular cloud cores by
  numerical simulations, taking into account turbulence and magnetic
  fields.  By using the adaptive mesh refinement technique,
  the collapse is followed over a wide dynamic range from the scale of a turbulent
  cloud core to that of the first core.
  The cloud core is lumpy in the low density region owing to the
  turbulence, while it has a smooth density distribution in the dense
  region produced by the collapse.
  The shape of the dense region depends mainly on the mass of the cloud
  core; a massive cloud core tends to be prolate while a less
  massive cloud core tends to be oblate.  In both cases, anisotropy of
  the dense region increases during the isothermal collapse ($n
  \lesssim 10^{11}\,\mathrm{cm}^{-3}$).  The minor axis of the dense
  region is always oriented parallel to the local magnetic field.
  All the models eventually yield spherical first cores ($n
  \gtrsim 10^{13}\,\mathrm{cm}^{-3}$)
  supported mainly by the thermal pressure.
  Most of turbulent cloud cores exhibit protostellar outflows around
  the first cores.  These outflows are classified into two types,
  bipolar and spiral flows, according to the morphology of the
  associated magnetic field.  Bipolar flow often appears in the
  less massive cloud core.  The rotation axis of the first core is
  oriented parallel to the local magnetic field for bipolar flow,
  while the orientation of the rotation axis from the global magnetic field depends on the
  magnetic field strength.   In spiral flow,
  the rotation axis is not aligned with the local magnetic
  field.

\end{abstract}


\section{Introduction}
Magnetic fields and interstellar turbulence are believed to play
important roles in the gravitational collapse of molecular cloud
cores.  
The measured magnetic fields of molecular clouds and molecular cloud cores
are strong and the magnetic energy is approximately equal to the
kinetic energy \citep{Crutcher99}.
Magnetic fields therefore have the potential to control 
the gravitational collapse of cloud cores.
Molecular clouds exhibit supersonic line widths, which are interpreted
as supersonic turbulence \citep{Zuckerman74}.  Such supersonic
turbulence seems to be common in the wide range from
molecular clouds to molecular cloud cores \citep{Langer95}.

% cloud shape: observations
One of the important properties of molecular cloud cores is their shape,
which is likely related to magnetic fields and turbulence \citep[see, e.g., the review by ][]{McKee07}.  Some studies of the
shape of cloud cores have suggested that they tend to be prolate
\citep{Myers91,Ryden96}.  The origin of the shape is however still
unknown.
% Outflow
Protostellar outflow is also an important feature related to magnetic
fields. Recent high-resolution observations of submillimeter
polarization succeeded in resolving the magnetic field around young stars to $\sim
10^{3-4}\,\mathrm{AU}$ \revise{\citep[e.g.,][]{Henning01,Wolf03,Valle03,Girart06}},
revealing the structure of the magnetic fields on such scales.  The outflows
often tend to be aligned with the magnetic field, while some are oriented perpendicular to it
\citep[][]{Wolf03}. Thus, there is as yet no clear correlation between
outflow and magnetic field.

% \citet{Jijina99}

% large scale simulations and turbulent simulation
There have been very few theoretical studies of collapse of magnetized
turbulent cloud cores in protostars, despite the importance of turbulence and magnetic fields.
Although self-gravitational turbulent simulations have been performed by many
researchers \citep[e.g.,][]{Gammie03,Li04,Offner08,Bate09}, most investigated large-scale turbulence and focused on cloud
core formation.
% they did not follow
% protostellar collapse to form the Larson's first
% core \citep{Larson69}.  
A notable exception is the work of \citet{Offner08}, who performed
high-resolution simulations using the adaptive mesh refinement (AMR)
technique to study protostellar collapse.  
However, these simulations did not include the effects of magnetic
fields. \citet{Goodwin04} also performed high-resolution investigations on protostellar collapse of
turbulent cores, but they also did not take magnetic fields into account.

% MHD protostellar collapse
On the other hand, simulations carried out to date on the formation of the Larson first core \citep{Larson69} have taken account of rotation and magnetic fields
but not turbulence,
both for aligned rotators
\citep{Tomisaka02,Machida04,Banerjee06,Commercon10,Tomida10} and
inclined rotators \citep{Matsumoto04,Machida06,Hennebelle09}.  In
these studies, the rotation speed and rotation axis are explicitly assumed 
as initial conditions, and the origin of the rotation has not been addressed.
\citet{Berkert00} showed that rotation derived from
observations could be reproduced by assuming the presence of turbulence with a power spectrum
of $P(k) \propto k^{-n}$ with $n = 3 -4$.  This suggests that the
rotation of cloud cores originates in turbulence.
Consequently, taking account of turbulence, we can incorporate rotation
in the simulations of protostellar collapse.  The present work
therefore investigates protostellar collapse of cloud cores,
considering both the turbulence and magnetic field.  
The AMR technique
is adopted in order to resolve the wide dynamic range from the cloud
core to the first core.

This paper is organized as follows.
In \S\ref{sec:model} and \S\ref{sec:methods}, the
model and simulation methods are presented.  The results of
the simulations are shown in \S\ref{sec:results}, and they are discussed
in \S\ref{sec:discussion}.  Finally, some conclusions are given in
\S\ref{sec:summary}.

\section{Models of Cloud Cores}
\label{sec:model}

As an initial model of a molecular cloud core, we consider a
turbulent, spherical cloud threaded by a uniform magnetic field. The
cloud is confined by a uniform ambient gas. 

As a template for a molecular cloud core, we consider the density
profile of the critical Bonner-Ebert sphere \citep{Bonnor1956,Ebert1955}.
When $\varrho_{\rm BE}(\xi)$ denotes the
non-dimensional density profile of the critical Bonnor-Ebert sphere
\citep[see][]{Chandrasekhar39}, the initial density distribution is
given by
\begin{equation}
\rho(r) = \left\{
\begin{array}{ll}
\rho_0 \varrho_{\rm BE}(r/a) & {\rm for}\; r < R_c\\ \rho_0
\varrho_{\rm BE}(R_c/a) & {\rm for}\; r \geq R_c
\end{array}
\right. \;,
\end{equation}
and
\begin{equation}
a = c_s \left( \frac{f}{4 \pi G \rho_0} \right)^{1/2} \;,
\label{eq:density-enhancement}
\end{equation}
where $r$, $G$, $c_s$, and $\rho_0$ denote the radius, gravitational
constant, isothermal sound speed, and initial central density,
respectively.  The gas temperature is assumed to be 10 K ($c_s =
0.19~\mathrm{km}~\mathrm{s}^{-1}$) .  
The initial central density is set equal to $\rho_0 = 10^{-19}\,{\rm g}\,{\rm
  cm}^{-3}$, which corresponds to a number density of $n_0 = 2.61
\times 10^4\,{\rm cm}^{-3}$ for an assumed mean molecular weight of
2.3.  
The non-dimensional parameter
$f$ denotes the density enhancement factor. 
The critical
Bonnor-Ebert sphere is obtained when $f=1$. 
An increase in density by a factor $f$ is equivalent to 
an enlargement of the spatial scale by a factor $f^{1/2}$ for a given central density.
The radius of the cloud is
defined by $R_c = 6.45 a = 0.137 f^{1/2}\,\mathrm{pc}$, 
where the numerical factor 6.45 comes from the
non-dimensional radius of the critical Bonnor-Ebert sphere. The
density contrast of the initial cloud is $\rho(0)/\rho(R_c) = 14.0$.
The initial freefall timescale at the center of the cloud is
thus $t_{\rm ff} \equiv (3 \pi / 32 G \rho_0)^{1/2} =
2.10\times10^5\,{\rm yr} $.
The mass of the cloud core ($r\le R_c$) is $M = 2.81 f^{3/2} M_\odot$.
The spherical cloud described above is located at the center of 
the computational domain of 
$x, y, z \in [-2R_c, 2R_c]\times [-2R_c, 2R_c]\times [-2R_c, 2R_c]$.

\revise{ Turbulence is given as the initial velocity field, and it is
  not driven in course of the simulations, because prestellar cloud
  cores are considered here and no driving source exists therein.
  Therefore, free decaying turbulence is considered.
}
The initial velocity field is incompressible with a power
spectrum of $P(k) \propto k^{-4}$, generated according to
\citet{Dubinski95}, where $k$ is the wavenumber.  This power spectrum
results in a velocity dispersion of $\sigma(\lambda) \propto
\lambda^{1/2}$, in agreement with the Larson scaling relations
\citep{Larson81}.
\revise{
  Note that the scaling relation is applied only at the initial
  condition, and the velocity dispersion changes during decay of
  the turbulence and collapse of the cloud cores.
}
The models are constructed by changing the mean
Mach number of the initial velocity field in the range ${\cal M} = 0 -
3$ with a common template of the initial velocity field.

The initial magnetic field is uniform in the $z$-direction.
The field strength is given by $B_z = \alpha B_\mathrm{cr} $,
 where $\alpha$ denotes the non-dimensional flux-to-mass ratio, 
and $B_\mathrm{cr}$
denotes the critical field
strength given by $B_\mathrm{cr} = 2 \pi G^{1/2} \Sigma$
\citep{Nakano78,Tomisaka88}.
The central column density $\Sigma$ is calculated by $\Sigma =
\int_{-R_c}^{R_c} \rho dz = 5.38 \rho_0 a$, where the integral is performed along
a line passing through the center of the cloud core.
In this paper, we examine the magnetically supercritical core ($\alpha < 1$).
The initial field strength is 
estimated as $B_z = 57.2 \alpha  f^{1/2} \,\mu$G by using the model
parameters $\alpha$ and $f$.
Note that the model parameter $\alpha$ is the inverse of the
dimensionless mass-to-flux ratio $\mu$ ($\alpha = 1/\mu$).

We construct 10 models by changing the three parameters $(\alpha, {\cal M}, f)$.
Using these parameters, the energies inside the cloud core ($r\le
R_c$) are calculated according to Appendix~\ref{seq:energies_of_be}.
The ratio of the thermal energy to the
gravitational energy is expressed as
$E_\mathrm{th}/|E_\mathrm{grav}| = 0.836 f^{-1}$.
When $f = 1.86$, 3.0, and 6.0, we obtain
$E_\mathrm{th}/|E_\mathrm{grav}| = 0.50$, 0.28, and 0.14, respectively.
The ratio of the kinetic energy to the gravitational energy is
expressed as $E_\mathrm{kin}/|E_\mathrm{grav}| = 0.394 f^{-1}{\cal M}^2$.
For example, 
the models with $({\cal M}, f) = (1.0, 1.68)$, $(3.0, 1.68)$, and $(3.0, 6.0)$
exhibit $E_\mathrm{kin}/|E_\mathrm{grav}| = 0.234$, 2.11 and 0.591, respectively.
Note that the numerical factor 0.394 depends on the seed of 
the random initial velocity field.
Finally, the ratio of the magnetic energy to the gravitational energy
is expressed as $E_\mathrm{mag}/|E_\mathrm{grav}| = 11.5\alpha^2$.
For example, the models with $\alpha = 0.1$ and 0.25 have 
$E_\mathrm{mag}/|E_\mathrm{grav}| = 0.115$ and 0.718, respectively.

The dynamical evolution of the cloud is followed by taking account of the
self-gravity, magnetic field, turbulence, and gas pressure.  The ideal magnetohydrodynamics  (MHD) is assumed here for simplicity.  The barotropic equation of state (EOS) is assumed where the gas temperature is $10$~K below the critical density
$\rho_\mathrm{cr} = 2 \times 10^{-13}\,\mathrm{g}~\mathrm{cm}^{-3}$ ($n_\mathrm{cr} =
5.24 \times 10^{10}~\mathrm{cm}^{-3}$), and it increases with
the adiabatic index $\gamma = 7/5$ above $\rho_\mathrm{cr}$.  This change in temperature
reproduces the formation of the adiabatic core, which corresponds to
the first core of \citet{Larson69}. The value of the critical density
$\rho_\mathrm{cr}$ is taken from the numerical results of \citet{Masunaga98},
who studied the spherical collapse of molecular cloud cores with
radiation hydrodynamics.  Recent multidimensional simulations using
radiation MHD produced results for protostellar collapse only quantitatively different
from simulations assuming a barotropic EOS \citep{Commercon10,Tomida10}.
The significant differences are restricted within the inside and proximity of
the first core.
% Radiation MHD simulations however require too huge computational
% cost to perform a parameter survey, which is a methodology adopted in
% the present work.

\section{Numerical Methods}
\label{sec:methods}
We calculated the evolution of the cloud cores using the AMR code, SFUMATO \citep{Matsumoto07}. 
It adopts a block-structured grid as the grid of the AMR hierarchy.
The total variation diminishing (TVD) cell-centered scheme is adopted
as the MHD solver, with the hyperbolic divergence cleaning method of
\citet{Dedner02}.
The MHD solver achieves second-order accuracy in space and time.
The self-gravity is solved by the multigrid method, exhibiting spatial second-order accuracy. 
The numerical fluxes are 
conserved by using a refluxing procedure in both the MHD and self-gravity solvers.
Periodic boundary conditions are imposed.

The computational domain of $[-2R_c, 2R_c]^3$ is initially resolved by a
uniform grid having $256^3$ cells.
The initial resolutions are 
$\Delta x = 2.8\times 10^{-3}\,\mathrm{pc}$,
$3.7\times 10^{-3}\,\mathrm{pc}$, and 
$5.3\times 10^{-3}\,\mathrm{pc}$ 
for $f = 1.68$, 3.0, and 6.0, respectively.
The Jeans condition is employed as a refinement criterion;
blocks are refined when the Jeans length is shorter than 8 times the cell width, i.e.,
$(\pi c_p^2/G\rho)^{1/2} < 8 \Delta x $ \citep[c.f.,][]{Truelove97},
where $c_p$ denotes the sound speed, and it is a function of density
in the barotropic EOS.  The finest resolution is
$7.0\times10^{-2}\,\mathrm{AU}$ for the typical case.

We followed the collapse beyond the stage in which 
the maximum density exceeds $\rho_\mathrm{max} = 10 ^{11}
\rho_0 = 10^{-8}\,\mathrm{g}\,\mathrm{cm}^{-3}$ ($n_\mathrm{max} =
2.62\times 10^{15}\,\mathrm{cm}^{-3}$) for all the models except for
the model with fast turbulence and a strong field where
$(\alpha, {\cal M}, f) = (0.5, 3.0, 1.68)$.
In this model, we confirmed that the cloud
does not undergo collapse even by  $t=15.5 t_\mathrm{ff}$ 
($ = 3.28\times10^{6}$~yr).

\section{Results}
\label{sec:results}
\subsection{Less massive cloud cores}
\label{sec:lowmass}
\subsubsection{Overview}
Less massive cloud cores, corresponding to models with $f = 1.68$, are examined
in this section.
All the models with $f = 1.68$ except for the model with 
$(\alpha, {\cal M}, f) = (0.5, 3.0, 1.68)$
undergo collapse.
After the collapse begins, the maximum density of the cloud core increases
rapidly. After it exceeds the critical density of the EOS
($\rho_\mathrm{cr}$), the cloud core forms an adiabatic core
supported against gravity mainly by the thermal pressure. The density of the adiabatic core is typically $\rho \gtrsim
10^{-10}\,\mathrm{g}\,\mathrm{cm}^{-3}$, which is approximately $10^3$
times larger than the critical density of the EOS.
The adiabatic core corresponds to the first core of
\citet{Larson69}, and hereafter we refer to it simply as the first core.
Models with larger $\alpha$ and/or ${\cal M}$ have longer latency
before the initiation of the collapse and hence 
first core formation.
The first core forms at $t = 2.39 t_\mathrm{ff}$ for 
a model with $(\alpha, {\cal M}, f) = (0.1, 1.0, 1.68)$, and at $t = 11.5
t_\mathrm{ff}$ for a model with $(\alpha, {\cal M}, f) = (0.25, 3.0, 1.68)$.

\begin{figure}
\epsscale{1.0}
\plotone{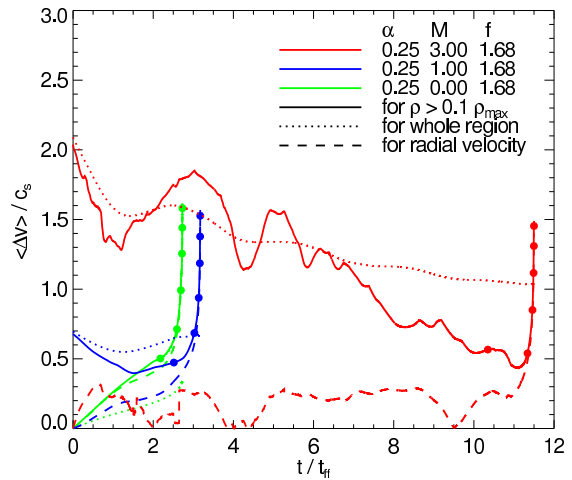}
% \plotone{vnt_rhomax2.eps}
\figcaption[vnt_time.eps]{ 
\revise{
  Velocity dispersion, $\left<\Delta v\right>$,
  as a function of time in the isothermal collapse phase ($\rho_\mathrm{max} \leq \rho_\mathrm{cr}$).
  Red, blue, and green lines correspond to models with 
  $(\alpha, {\cal M}, f) = (0.25, 3.0, 1.68)$, 
  (0.25, 1.0, 1.68), and (0.25, 0.0, 1.68), respectively. 
  Solid lines denote
  velocity dispersion in the dense
  region of $\rho \ge 0.1 \rho_\mathrm{max}$, while the dotted lines
  denote velocity dispersion in the whole computational domain.
  Dashed lines is same as the solid lines but for the radial velocity,
  $\left<\Delta v_r\right>$.
  Filled circles associated with the solid lines denote the stages of 
  $\rho_\mathrm{max} = 10^n \rho_0$ ($n = 1, 2, \cdots, 6$).  
}
  \label{vnt_time.eps}
}
\end{figure}

\revise{ 
  Figure~\ref{vnt_time.eps} shows the evolution of the velocity
  dispersions in the isothermal collapse phase ($\rho_\mathrm{max} \le
  \rho_\mathrm{cr}$) for models with moderate magnetic fields ($\alpha
  = 0.25$).  The velocity dispersions within the dense region of
  $\rho \ge 0.1\rho_\mathrm{max}$ are calculated according to
  equation~(\ref{eq:delta_v}).  The velocity dispersions decrease in the
  dense region before the collapse begins for the turbulent
  models (solid lines).  When the collapse sets in, the velocity
  dispersion is subsonic even for the strong turbulent model (${\cal M} =
  3.0$) as denoted by the red solid line.  In other words, decay of the
  turbulence promotes the collapse in the dense region.  This is
  consistent with molecular line observations of the dense cores where
  narrow line widths are obtained.
  The velocity dispersion is smaller in the dense region than in the
  whole computational domain (dotted lines) when collapse sets in.
  This indicates that the turbulence decays in the collapsing dense
  region selectively, and the other region remains turbulent even
  after the collapse sets in.  }

\revise{ 
  As the collapse proceeds, the velocity dispersion increases and it
  exceeds the sound speed in the dense region.  The increase in the
  velocity dispersion is attributed to the infall motion as denoted by
  the dashed lines (see eq.~[\ref{eq:delta_vr}]).  The radial infall
  dominates over the velocity dispersion in the stages of
  $\rho_\mathrm{max} \gtrsim 10^{-17}\,\mathrm{g}\,\mathrm{cm}^{-3}$.
  Note that the initial stage exhibits $\left<\Delta v \right>/c_s
  \simeq 2 $ for the model with ${\cal M} = 3.0$, and $\left<\Delta v
  \right>/c_s \simeq 0.7$ for the model with ${\cal M} = 1.0$.  These
  reductions of $\left<\Delta v \right>$ are caused by a density
  weighted average in the calculation of $\left<\Delta v \right>$.
}

\begin{figure}
\epsscale{1.0}
\plotone{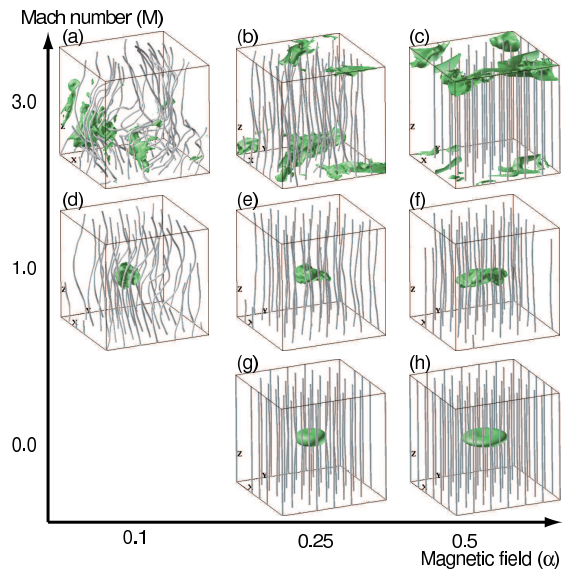}
\figcaption[avsWhole.eps]{ 
  Three-dimensional structures of density and magnetic fields for models
  with $f = 1.68$.  The stage
  $\rho_\mathrm{max} = 10^{-8}\,\mathrm{g}\,\mathrm{cm}^{-3}$ 
  ($n_\mathrm{max} = 2.62\times 10^{15}\,\mathrm{cm}^{-3}$)
  is shown for the collapse models,
  while the final stage of $t = 15.5 t_\mathrm{ff}$ is shown for 
  the non-collapse model of $(\alpha, {\cal M}, f) = (0.5, 3.0,
  1.68)$.
  Isosurfaces denote the iso-density
  surfaces of $\rho = 3.16 \times 10^{-20}\,\mathrm{g}\,\mathrm{cm}^{-3}$
  ($n = 8.28 \times 10^3 \,\mathrm{cm}^{-3}$). 
  Tubes denote magnetic field lines.
  The boxes enclose the entire computational domain $(0.712~\mathrm{pc})^3$.
\label{avsWhole.eps}
}
\end{figure}

Figure~\ref{avsWhole.eps} shows the cloud structures 
on the scale of the whole computational domain, $(0.712\,\mathrm{pc})^3$, 
at the stage with $\rho_\mathrm{max} = 10^{-8}\,\mathrm{g}\,\mathrm{cm}^{-3}$ for
the collapse models, and at $t = 15.5 t_\mathrm{ff}$ for
the non-collapse model with $(\alpha, {\cal M}, f) = (0.5, 3.0,
  1.68)$. 　
  The iso-density surfaces indicate the structures of the cloud cores
  at the low density of $\rho = 3.16 \times 10^{-20}\,\mathrm{g}\,\mathrm{cm}^{-3}$
  ($n = 8.28 \times 10^3 \,\mathrm{cm}^{-3}$),
  corresponding to the boundary between the cloud core and the
  parent cloud.
The models with a high Mach number exhibit
a density structure highly disturbed by the turbulence
(Fig.~\ref{avsWhole.eps}{\it a}--\ref{avsWhole.eps}{\it c}), where
the initial configuration of the cloud core disappears.
In models with a moderate Mach number, the cloud core is strongly disturbed
(Fig.~\ref{avsWhole.eps}{\it d}--\ref{avsWhole.eps}{\it f}). 
The models shown in Figure~\ref{avsWhole.eps}{\it e} and \ref{avsWhole.eps}{\it f}
produce cores that are flattened in the plane perpendicular to the magnetic field.
The models without turbulence produce axisymmetric oblate cores flattened
in the plane perpendicular to the magnetic field
(Fig.~\ref{avsWhole.eps}{\it g}--\ref{avsWhole.eps}{\it h}). 
The magnetic field lines are highly disturbed by the turbulence in
the weak-field models 
(Fig.~\ref{avsWhole.eps}{\it a} and \ref{avsWhole.eps}{\it d}). 
In contrast, the strong-field models exhibit almost straight field lines
(Fig.~\ref{avsWhole.eps}{\it c}, \ref{avsWhole.eps}{\it f}, and \ref{avsWhole.eps}{\it h}).

\begin{figure}
\epsscale{1.0}
\plotone{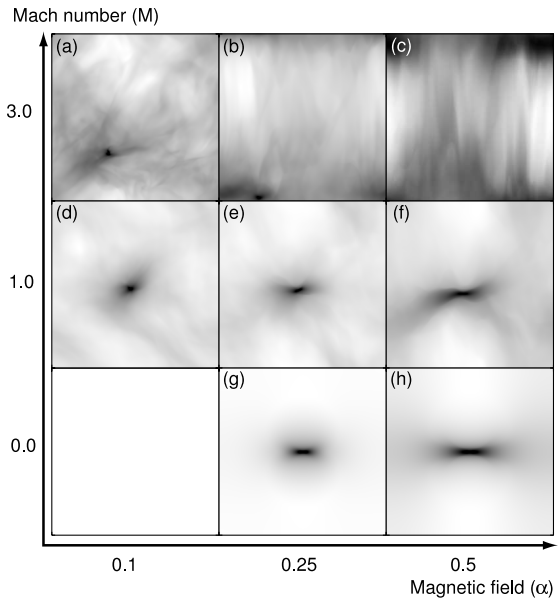}
\figcaption[whole.eps]{ 
Column density distributions of the entire computational domain
$(0.712~\mathrm{pc})^3$
at $\rho_\mathrm{max} = 10^{-8}\,\mathrm{g}\,\mathrm{cm}^{-3}$ 
($n_\mathrm{max} = 2.62\times 10^{15}\,\mathrm{cm}^{-3}$)
for the
collapse models with $f = 1.68$.  For the non-collapse model of
$(\alpha, {\cal M}, f) = (0.5, 3.0, 1.68)$, the final stage of $t =
3.28\times10^{6}$~yr is shown.
The column densities are shown on a logarithmic scale in the $x-z$
planes.
\label{whole.eps}
}
\end{figure}

Figure~\ref{whole.eps} shows the column density distributions of
the whole computational domain for the models shown in
Figure~\ref{avsWhole.eps}, clarifying the fine density distribution.
The cloud cores are shown on a density
scale of $n \simeq 10^4 \, \mathrm{cm}^{-3}$, corresponding to
the C$^{18}$O cores.
In the model with a high Mach number and weak field (Fig.~\ref{whole.eps}{\it a}), 
the low density region is highly disturbed by the turbulence,
and exhibits a complex structure. 
The model with a high Mach number and strong magnetic field
(Fig.~\ref{whole.eps}{\it c}) forms a molecular gas sheet, which lies
at the top boundary of the computational domain.  
The location of the sheet depends on the seed of 
the random initial velocity field.
This model shows a filamentary structure aligned parallel to the magnetic
field.  A similar structure was reported by \citet{Price08}, who
performed an SPH simulation taking account of the magnetic field.
\revise{\citet{Nakamura08} also reported such a filamentary structure
  by performing grid based MHD simulations including ambipolar diffusion.}
Oblate cloud cores form in Figures~\ref{whole.eps}{\it e},
\ref{whole.eps}{\it f}, \ref{whole.eps}{\it g}, and
\ref{whole.eps}{\it h}.
The cloud cores are more flattened when the magnetic field is strong.
In the turbulent models, the edges of the oblate cloud cores are warped.

At the stage shown in Figure~\ref{whole.eps}, all the turbulent cloud cores move at
subsonic speeds. When we define the cloud core as the gas denser
than the initial ambient gas ($\rho > \rho_0/14$), the velocities of
the baricenter of turbulent cloud cores range from
0.01 to $0.04\,\mathrm{km}\,\mathrm{s}^{-1}$.
Among all the models, the model of $(\alpha, {\cal M}, f) = (0.25, 3.0, 
1.68)$ produces the first core at the most distant point (0.39~pc) from
the initial peak of the cloud core.

\begin{figure}
\epsscale{1.0}
\plotone{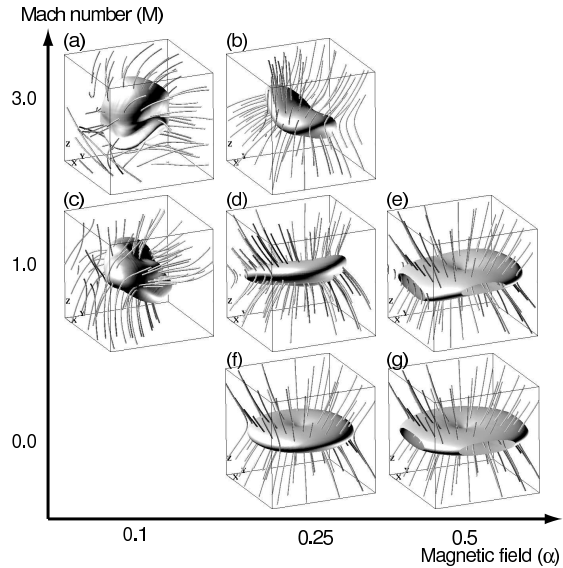}
\figcaption[avs2000au.eps]{ 
Same as Figure~\ref{avsWhole.eps} but for a $(4000~\mathrm{AU})^3$ region.
Isosurfaces denote the iso-density surfaces of $\rho = 10^{-17}\,\mathrm{g}\,\mathrm{cm}^{-3} $
($n = 2.62 \times 10^6 \,\mathrm{cm}^{-3}$). 
The model with $(\alpha, {\cal M},f ) = (0.5, 3.0, 1.68)$ is not shown because it
does not undergo collapse and the maximum density is lower than the level 
of the iso-surface.
\label{avs2000au.eps}
}
\end{figure}

Figure~\ref{avs2000au.eps} shows the density and magnetic
field in the dense regions of $(4000~\mathrm{AU})^3$ for the collapse
models, showing iso-density surfaces of $\rho =
10^{-17}\,\mathrm{g}\,\mathrm{cm}^{-3}$ ($n = 2.62 \times 10^6 \,\mathrm{cm}^{-3}$). 
The cloud cores shown in this figure correspond to the dense cores
observed by dust continuum emissions.
Even on this scale, the weak-field models (Figs.~\ref{avs2000au.eps}{\it a}
and \ref{avs2000au.eps}{\it c})
 exhibit a disturbed density structure, and 
the magnetic field lines are not aligned. 
The remaining models exhibit oblate shapes with a minor axis parallel to the local mean magnetic field.  The magnetic field lines are 
 in the configuration of an hourglass.
The direction of the mean magnetic field depends on the Mach number
and the magnetic field strength.
The direction of the local magnetic field in models with a higher Mach number and
weaker magnetic field is inclined more from that of the initial (global)
magnetic field.
The models of Figures~\ref{avs2000au.eps}{\it e}, \ref{avs2000au.eps}{\it f}, and
\ref{avs2000au.eps}{\it g} exhibit mean magnetic fields parallel
to the $z$-direction, while the models of
Figures~\ref{avs2000au.eps}{\it b}, \ref{avs2000au.eps}{\it c}, and
\ref{avs2000au.eps}{\it d} exhibit mean magnetic fields oriented in other directions.

\begin{figure}
\epsscale{1.0}
\plotone{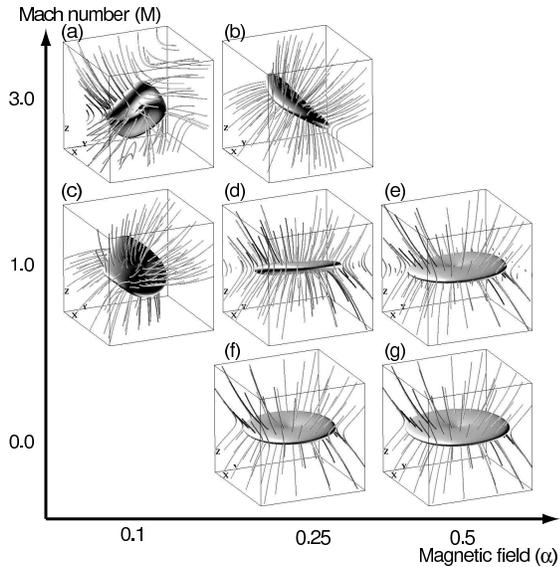}
\figcaption[avs200au.eps]{ 
Same as Figure~\ref{avsWhole.eps} but for a $(400~\mathrm{AU})^3$ region.
Isosurfaces denote the iso-density surfaces of 
$\rho = 3.16 \times 10^{-15}\,\mathrm{g}\,\mathrm{cm}^{-3}$
($n =  8.28 \times 10^8 \,\mathrm{cm}^{-3}$). 
\label{avs200au.eps}
}
\end{figure}

Figure~\ref{avs200au.eps} shows the density and the magnetic
field on a 400~AU scale, showing iso-density surfaces
of $\rho = 3.16 \times 10^{-15}\,\mathrm{g}\,\mathrm{cm}^{-3}$
($n =  8.28 \times 10^8 \,\mathrm{cm}^{-3}$).
On this scale, all the models except for $(\alpha, {\cal M}, f) = (0.1,
3.0, 1.68)$ exhibit flat disks in the plane perpendicular to the hourglass-shaped magnetic fields. 
The model with $(\alpha, {\cal M}, f) = (0.1, 3.0, 1.68)$ produces a highly warped disk, and the magnetic field lines are not aligned even
on this scale.

\begin{figure}
\epsscale{1.0}
\plotone{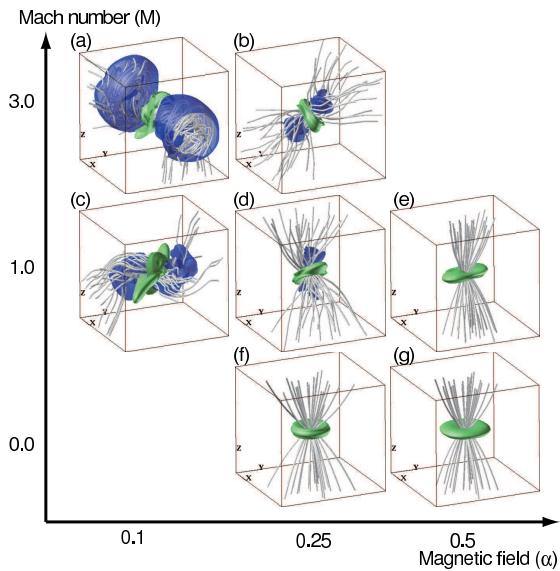}
\figcaption[Outflows_allmodel.eps]{ 
Densities, magnetic fields, and outflows at the outflow formation
stages for the model with $f = 1.68$ in the dense
region with $(40~\mathrm{AU})^3$.
The stages are 
$\rho_\mathrm{max} = $
({\it a}) $1.08\times 10^{-8}\,\mathrm{g}\,\mathrm{cm}^{-3}$,
({\it b}) $1.01\times 10^{-8}\,\mathrm{g}\,\mathrm{cm}^{-3}$,
({\it c}) $7.55\times 10^{-8}\,\mathrm{g}\,\mathrm{cm}^{-3}$,
({\it d}) $1.27\times 10^{-8}\,\mathrm{g}\,\mathrm{cm}^{-3}$,
({\it e}) $2.16\times 10^{-8}\,\mathrm{g}\,\mathrm{cm}^{-3}$,
({\it f}) $1.43\times 10^{-8}\,\mathrm{g}\,\mathrm{cm}^{-3}$,
({\it g}) $1.04\times 10^{-8}\,\mathrm{g}\,\mathrm{cm}^{-3}$.
The green isosurface represents a density of 
  $\rho = 1.00\times 10^{-12}\,\mathrm{g}\,\mathrm{cm}^{-3}$
  ($n = 2.62 \times 10^{11} \,\mathrm{cm}^{-3}$). 
The blue isosurface is for the radial velocity of 
$v_r = 0.57\,\mathrm{km}\,\mathrm{s}^{-1}$
($v_r = 3 c_s$).
The tubes indicate the magnetic field lines.
\label{Outflows_allmodel.eps}
}
\end{figure}

Figure~\ref{Outflows_allmodel.eps} shows the dense region on a
40~AU scale for the outflow formation stage.
Four of the five turbulent collapse models produce outflows
indicated by the blue isosurface of the radial
velocity of $v_r = 0.57\,\mathrm{km}\,\mathrm{s}^{-1}$
($v_r = 3 c_s$). The radial velocity is measured from the location
of maximum density.  
The outflow appears $\sim 200$~yr after the maximum density exceeds
the critical density of the EOS for all the outflow models.
This epoch corresponds to the stages with $\rho_\mathrm{max} \sim
10^{-9}\,\mathrm{g}\,\mathrm{cm}^{-3}$ for the model of
Figure~\ref{Outflows_allmodel.eps}{\it a}, and  
$\rho_\mathrm{max}  \sim 10^{-8}\,\mathrm{g}\,\mathrm{cm}^{-3}$ for 
the remaining outflow models.
All the outflows are ejected in the direction parallel to the local magnetic field.
The weak-field models 
(Figs.~\ref{Outflows_allmodel.eps}{\it a} and \ref{Outflows_allmodel.eps}{\it c})
result in outflow directions completely different from the
initial direction of the magnetic field. 
This is consistent with \citet{Matsumoto04}, who examined
outflow formation in collapsing clouds, in which 
the initial magnetic field and rotation axis were not aligned. 
The outflows reproduced here are classified into two types:
bipolar and spiral flows.  
The outflow shown in Figure~\ref{Outflows_allmodel.eps}{\it a} is
 a bipolar flow driven by tightly twisted magnetic fields.
The bipolar flows in Figure~\ref{Outflows_allmodel.eps}{\it b} and
\ref{Outflows_allmodel.eps}{\it d} are accelerated by magnetic fields that are less twisted
because of their higher strength. 
In these bipolar flows, the rotation axis of the central first
core is parallel to the mean direction of the local magnetic field.
The other weak-field model of Figure~\ref{Outflows_allmodel.eps}{\it
  c} shows a spiral flow, which is qualitatively different from the previous three
outflows, showing magnetic field lines wound in a spiral shape, along which the gas is accelerated. 
The rotation axis of the central first core is inclined at a
large angle of 35\,$^\circ$ to the direction of the local mean magnetic field.
The model shown in Figure~\ref{Outflows_allmodel.eps}{\it e} does not produce an
outflow even at the final stage with $\rho = 2.16 \times 10^{-8}\,\mathrm{g}\,\mathrm{cm}^{-3}$.  Because of the high magnetic field strength in this model, angular momentum is transferred mainly by magnetic braking instead of by outflow.
The relationship between the outflow, rotation, and magnetic field
are described in detail in \S\ref{sec:outflow}.

Note that we reproduce the very early phase of outflow formation.  At
the stages shown in Figure~\ref{Outflows_allmodel.eps}, the highest outflow velocity is found for Figure~\ref{Outflows_allmodel.eps}{\it a}, exhibiting a maximum value of
$v_r = 20.7 c_s$ ($= 3.9 \,\mathrm{km}\,\mathrm{s}^{-1}$). The remaining models produce maximum outflow velocities of $\sim 10 c_s$ ($= 2
\,\mathrm{km}\,\mathrm{s}^{-1}$).  For all models,  the outflow velocity continues to
increase until the end of the calculations.

\begin{figure}
\epsscale{1.0}
\plotone{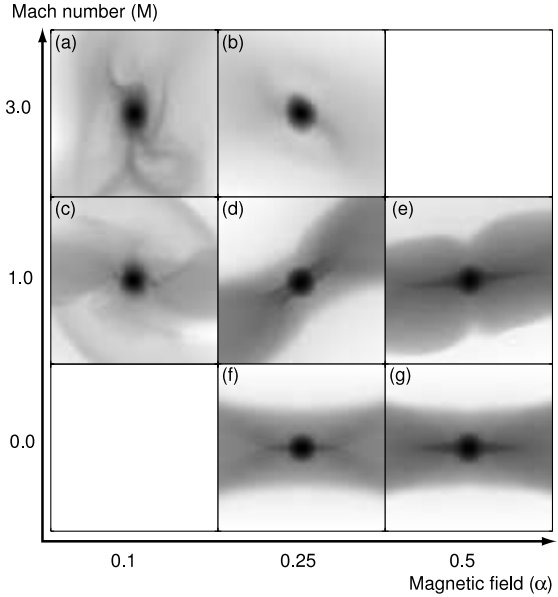}
\figcaption[central5AU.eps]{ 
Column density distributions of central $(10~\mathrm{AU})^2$ region 
for $\rho_\mathrm{max} = 10^{-8}\,\mathrm{g}\,\mathrm{cm}^{-3}$.
for the models with $f = 1.68$.
The column densities are shown on a logarithmic scale in the $x-z$
planes.
\label{central5AU.eps}
}
\end{figure}

Figure~\ref{central5AU.eps} shows close-up views of the first cores in 
the column density distributions on a 10~AU scale.
All the collapse models produce spherical first cores with a
radius of approximately $1$~AU.  
The masses of the first cores are $\sim 6\times10^{-3}\,M_\odot$ at
the stage shown in Figure~\ref{central5AU.eps}.
All the first cores are surrounded
by disk-shaped envelopes.  The disk shape can be clearly seen in 
Figures~\ref{central5AU.eps}{\it d}, \ref{central5AU.eps}{\it e},
\ref{central5AU.eps}{\it f}, and \ref{central5AU.eps}{\it g} because
these disks are being viewed edge-on.  
The outflows disturb the disk-shaped envelope near the first cores
(Fig.~\ref{central5AU.eps}{\it a} and \ref{central5AU.eps}{\it c}).
Note that both the first cores and the disk-shaped envelopes rotate slowly. 
The first cores are supported against gravity by their thermal pressure. The disk-shaped envelopes, on the other hand, are only partially supported by the centrifugal force;
consequently, the infall velocity dominates over  the rotation velocity.
The disk-shaped envelope resembles the magnetized pseudo
disk reported by \citet{Shu97}.  Recently, \citet{Mellon08,Mellon09} also reported
that a centrifugally supported disk cannot be formed in the presence of a magnetic field due to magnetic braking.

\subsubsection{Change in shape during collapse}

\begin{figure}
\epsscale{1.0}
\plotone{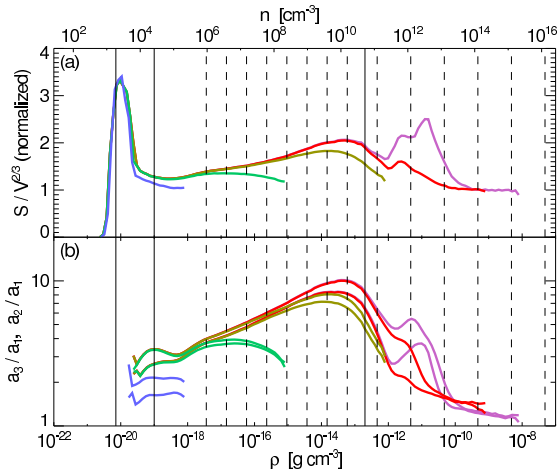}
\figcaption[surface_volume_M1B025.eps]{ 
({\it a}) Normalized surface-to-volume ratios and 
({\it b}) axis ratios as functions of
density thresholds
for the model with $(\alpha, {\cal M}, f ) = (0.25, 1.0, 1.68)$.
The different colored lines correspond to the stages
of $\rho_\mathrm{max}/\rho_0 = 10, 10^4, 10^7, 10^{10}$, and $10^{11}$
($\rho_\mathrm{max} = 10^{-18}, 10^{-15}, 10^{-12}, 10^{-9}, 10^{-8}\,\mathrm{g}\,\mathrm{cm}^{-3}$).
From left to right, the solid vertical lines represent the initial density of the ambient gas
($\rho_0/14.0$), the initial central density, and the critical density of the EOS
($\rho_\mathrm{cr}$).  
The dashed vertical lines represent
the densities at which grid refinement was carried out according to the Jeans condition.
\label{surface_volume_M1B025.eps}
}
\end{figure}

In order to measure the complexity of the density distribution during the
collapse, we evaluate the surface-to-volume ratio and axis ratios for a given
iso-density surface according to
Appendix~\ref{sec:surface-to-volume}.
The surface-to-volume ratio is normalized so that it has a value of unity for a
sphere.
Figure~\ref{surface_volume_M1B025.eps} shows surface-to-volume ratios
for the model with $(\alpha, {\cal M}, f ) = (0.25, 1.0, 1.68)$ as a fiducial model.
In the early stage with $\rho_\mathrm{max} = 10^{-18}\,\mathrm{g}\,\mathrm{cm}^{-3}$ (blue curves), 
the iso-surfaces of $\rho\simeq 10^{-20}\,\mathrm{g}\,\mathrm{cm}^{-3}$ have
high surface-to-volume ratios, indicating that the cloud core has
a complex density distribution in the low density region. 
\revise{
The surface-to-volume ratio vanishes below the minimum density.  This
is because the isosurface vanishes there while the volume coincide with that
of the whole computation box because of the periodic boundary
condition. }
In contrast, the iso-surface of $\rho \sim
10^{-18}\,\mathrm{g}\,\mathrm{cm}^{-3}$ exhibits a low 
surface-to-volume ratio, and $a_3$ is considerably longer than $a_2$.
This indicates that the collapsing region has a smooth triaxial shape
at this stage.
Note that the vertical dashed lines denote the densities at which the grid
refinements are performed.  At low densities, the surface-to-volume ratio is seen to make a transition from a high to a low value. Since this occurs at a density less than that of the first refinement, it cannot be attributed to the grid refinement process.
At the stage with $\rho_\mathrm{max} =
10^{-15}\,\mathrm{g}\,\mathrm{cm}^{-3}$ (green curves), the two axis
ratios have an identical value of $3-4$, indicating that the
collapsing region of the cloud core is an almost
axisymmetric disk.  Owing to its flat shape, the surface-to-volume
ratio is slightly higher than unity for $\rho \sim
10^{-17}\,\mathrm{g}\,\mathrm{cm}^{-3}$.
At the stage with $\rho_\mathrm{max} = 10^{-12}\,\mathrm{g}\,\mathrm{cm}^{-3}$ (yellow curves), 
the surface-to-volume ratio reaches 1.8 at $\rho \simeq 10^{-14}\,\mathrm{g}\,\mathrm{cm}^{-3}$, 
and the axis ratios reach $7-8$, indicating that the flatness of the
disk increases during the collapse.
At the stage with $\rho_\mathrm{max} = 10^{-9}\,\mathrm{g}\,\mathrm{cm}^{-3}$ (red curves), 
a spherical first core forms, and the
surface-to-volume ratio and axis ratios become unity 
at the high density of $\rho \gtrsim 10^{-10}\,\mathrm{g}\,\mathrm{cm}^{-3}$.
The surface-to-volume ratio and the axis ratios of the disk-shaped envelope
reach 2 and 10, respectively, at $\rho \simeq
10^{-13}\,\mathrm{g}\,\mathrm{cm}^{-3}$ owing to an increase in the flatness. 
At the outflow formation stages (purple curves), the surface-to-volume
ratio takes a high value of 2.5 at $\rho \simeq
10^{-11}\,\mathrm{g}\,\mathrm{cm}^{-3}$, reflecting the fact that the outflow
disturbs the disk-shaped envelope around the first core.
The  surface-to-volume ratio and the axis ratios remain unity in 
the dense region with $\rho \gtrsim 10^{-10}\,\mathrm{g}\,\mathrm{cm}^{-3}$,
indicating that the first core is spherical even at the
outflow formation stage.
Moreover, at the low density region with $\rho \simeq
10^{-20}\,\mathrm{g}\,\mathrm{cm}^{-3}$, the surface-to-volume ratio
remains high even after first core formation, indicating that the
periphery of the cloud core remains turbulent.

\begin{figure}
\epsscale{1.0}
\plotone{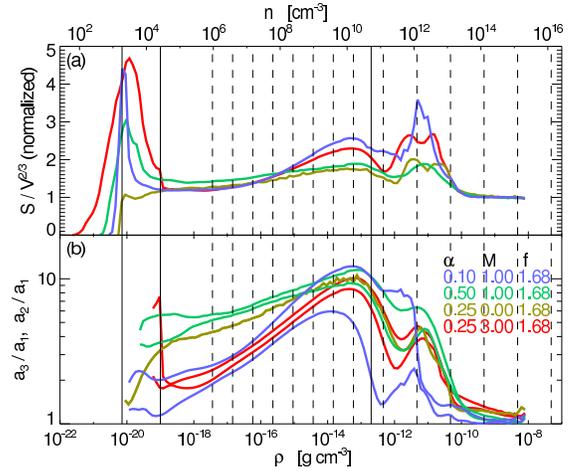}
\figcaption[surface_volume_allmodels.eps]{ 
({\it a}) Normalized surface-to-volume ratios and 
({\it b}) axis ratios as functions of
density thresholds
for the models with $f =
1.68$ at the stage with $\rho_\mathrm{max} = 10^{-8}\,\mathrm{g}\,\mathrm{cm}^{-3}$. 
The different colored lines correspond to the different models.
From left to right, the solid vertical lines represent the initial density of the ambient gas
($\rho_0/14.0$), the initial central density, and the critical density of the EOS
($\rho_\mathrm{cr}$).  
The dashed vertical lines represent
the densities at which grid refinement was carried out according to the Jeans condition.
\label{surface_volume_allmodels.eps}
}
\end{figure}

Figure~\ref{surface_volume_allmodels.eps} compares the surface-to-volume ratio and the axis ratios for the four models with $f = 1.68$.
For a low density of
$\rho \lesssim 10^{-19}\,\mathrm{g}\,\mathrm{cm}^{-3}$, 
the turbulent models exhibit high
surface-to-volume ratios.  The model with a high Mach
number (${\cal M}=3$) exhibits the highest value (red curve), while the
model without turbulence exhibits a low ratio (yellow  curve).
Comparing the models with ${\cal M} = 1$, the weak-field model (blue
curve) has a higher surface-to-volume ratio than the strong-field model
(green curve), implying that disturbance by turbulent flow is
considerably suppressed by the magnetic field. 

In all the models, the axis ratios increase with density 
in the range of $10^{-18}\,\mathrm{g}\,\mathrm{cm}^{-3} \lesssim \rho \lesssim 10^{-13}\,\mathrm{g}\,\mathrm{cm}^{-3}$.
In this range, the mean axis ratios $(a_2+a_3)/(2a_1)$ tend to increase
with the initial magnetic filed strength $\alpha$, and decrease with 
the initial Mach number ${\cal M}$.  
This implies that the magnetic
field increases the degree of anisotropy and the turbulence increases the effective sound speed.
This tendency is apparent at a relatively low density of
$10^{-18}\,\mathrm{g}\,\mathrm{cm}^{-3} \lesssim \rho \lesssim 10^{-15}\,\mathrm{g}\,\mathrm{cm}^{-3}$.

At $\rho \sim 10^{-13}\,\mathrm{g}\,\mathrm{cm}^{-3}$, 
the model with $(\alpha, {\cal M}, f) = (0.1, 1.0, 1.68)$ exhibits a
prolate shape (blue curves), 
with the major axis being considerably longer than the other axes; the axis ratio $a_3/a_2$ has a maximum of 6.2 at $\rho = 3.8\times 10^{-13}\,\mathrm{g}\,\mathrm{cm}^{-3}$. 
For all the other models, the axes $a_3$ and $a_2$ are comparable ($a_3/a_2 < 2$) for $\rho \ge 10^{-17}\,\mathrm{g}\,\mathrm{cm}^{-3}$, 
indicating an oblate shape.
For a higher density of $\rho \gtrsim 10^{-11}\,\mathrm{g}\,\mathrm{cm}^{-3}$, 
all the models exhibit spherical shapes because of the presence of the first cores. 

\subsection{Dependence on mass}
\label{sec:highmass}

We examined two additional models with $f = 3.0$ and 6.0 in order to
investigate the dependence on cloud mass.  We refer to these models
as massive cloud cores.
The parameters of the initial magnetic field and the initial Mach
number of turbulence are $\alpha = 0.1$ and ${\cal M} = 3.0$, respectively.
\revise{
It takes shorter time to form first core for the more massive core; 
the first core forms at $t = 5.24 t_\mathrm{ff}$ for a model with $(\alpha, {\cal
  M}, f) = (0.1, 3.0, 1.68)$, and 
at $t = 1.29 t_\mathrm{ff}$ for a model with $(\alpha, {\cal
  M}, f) = (0.1, 3.0, 6.0)$.
}

\begin{figure}
\epsscale{1.0}
\plotone{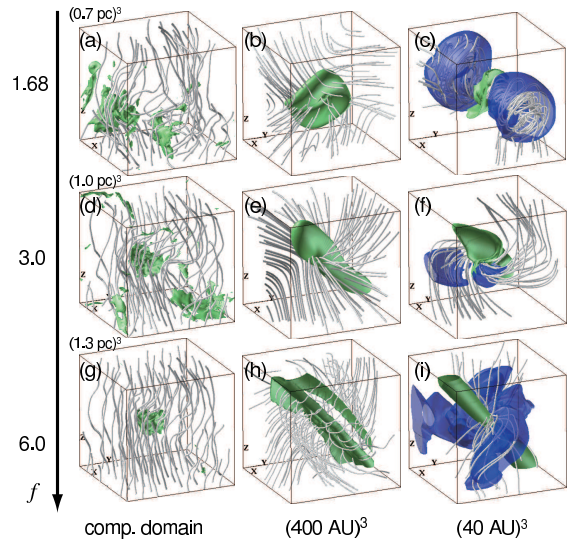}
\figcaption[allmodel_F.eps]{ 
  Three-dimensional structures of density, magnetic fields, and
  outflows for the models
  with 
  \revise{
  $(\alpha, {\cal M}, f) = (0.1, 3.0, 1.68)$, $(0.1, 3.0, 3.0)$, and
  $(0.1, 3.0, 6.0)$}
  from top to bottom.
  From left to right, the plotted areas are the entire computation box,  $(400~\mathrm{AU})^3$,
  and $(40~\mathrm{AU})^3$. 
  From left to right, the isosurfaces denote the iso-density surfaces of 
  $\rho = 3.16 \times 10^{-20}\,\mathrm{g}\,\mathrm{cm}^{-3}$, 
  $3.16 \times 10^{-15}\,\mathrm{g}\,\mathrm{cm}^{-3}$, and
  $1.00\times 10^{-12}\,\mathrm{g}\,\mathrm{cm}^{-3}$.
  Tubes denote magnetic field lines.
  The blue isosurface is for a radial velocity of 
  $v_r = 0.57\,\mathrm{km}\,\mathrm{s}^{-1}$
  ($v_r = 3 c_s$).
  The stages are $\rho_\mathrm{max} =
  4.53\times10^{-8}\,\mathrm{g}\,\mathrm{cm}^{-3}$ for panel {\it f},
  and $\rho_\mathrm{max} = 10^{-8}\,\mathrm{g}\,\mathrm{cm}^{-3}$ for
  the rest of the panels.
\label{allmodel_F.eps}
}
\end{figure}

\begin{figure}
\epsscale{1.0}
\plotone{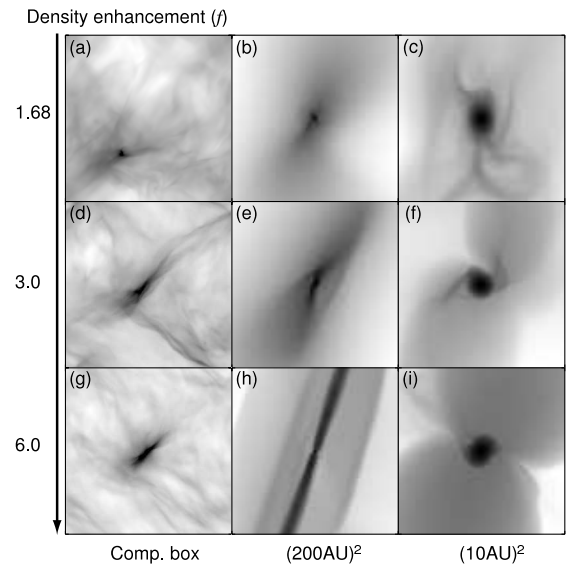}
\figcaption[wholeF.eps]{ 
Column density distributions at the stage with $\rho_\mathrm{max} = 10^{-8}\,\mathrm{g}\,\mathrm{cm}^{-3}$ 
for the models with 
\revise{
$(\alpha, {\cal M}, f) = (0.1, 3.0, 1.68)$, $(0.1, 3.0, 3.0)$, and
  $(0.1, 3.0, 6.0)$}
from top to bottom.
The column densities are shown on a logarithmic scale in the $x-z$
planes.  From left to right, the plotted regions are
the whole computational domains, $(200~\mathrm{AU})^2$, $(10~\mathrm{AU})^2$.
\label{wholeF.eps}
}
\end{figure}

Figures~\ref{allmodel_F.eps} and \ref{wholeF.eps} compare the three
models on three spatial scales.  
The low density regions are highly disturbed by the turbulence 
in all the models (left column in
Figs.~\ref{allmodel_F.eps} and \ref{wholeF.eps}). On the intermediate
scale (the middle column), the clouds take the shape of filaments for the
massive models.  The most massive model produces a long thin filament as
shown in Figures~\ref{allmodel_F.eps}{\it h} and \ref{wholeF.eps}{\it h}.
On the small scale, all the models produce spherical first cores (right column in Fig.~\ref{wholeF.eps}).  
The first cores of the massive clouds are embedded in the filament, while that of
the less massive cloud is surrounded by the disk (right column in
Figs.~\ref{allmodel_F.eps}). 

All the models exhibit outflows as shown by the blue iso-velocity surfaces in
Figure~\ref{allmodel_F.eps}, but their shapes are different.  
The outflow shown in Figure~\ref{allmodel_F.eps}{\it c} is 
classified as bipolar flow, and it is ejected in the direction
perpendicular to the plane of the disk.
The disk-outflow system is roughly
axisymmetric although its axis is inclined considerably with respect
to the initial direction of the magnetic field (the vertical direction).  
In the massive models with $f = 3.0$ and 6.0
(Figs.~\ref{allmodel_F.eps}{\it f} and \ref{allmodel_F.eps}{\it i}), 
the outflow is classified as spiral flow, and it is ejected in the plane perpendicular
to the filament, and is not collimated.
Similar outflow appears for the model with $(\alpha, {\cal M}, f)
= (0.1, 1.0, 1.68)$ as shown in Figure~\ref{Outflows_allmodel.eps}{\it
c}.
The configuration of the magnetic field depends on the type of outflow present.  In
the less massive model with $f=1.68$, the magnetic field lines are
twisted and highly wound along the bipolar directions, implying
that the outflow is accelerated by the magnetic pressure.  In the
massive models, the magnetic field lines exhibit a spiral configuration, which is attributed to misalignment between the rotation axis
and the local mean magnetic field.  This morphology implies
that the outflow is accelerated by magnetic tension.

%% Surface-volume ratio and axis ratio

\begin{figure}
\epsscale{1.0}
\plotone{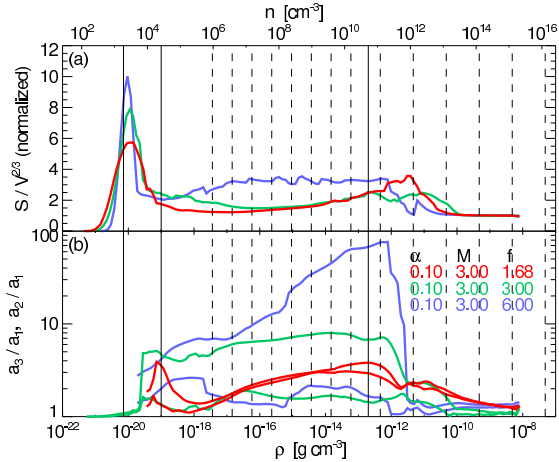}
\figcaption[surface_volume_allmodels_f.eps]{ 
({\it a}) Normalized surface-to-volume ratios and 
({\it b}) axis ratios as functions of
density thresholds
for the models with ${\cal M} =3$ and $\alpha = 0.1$. 
From left to right, the solid vertical lines represent the initial density of the ambient gas
($\rho_0/14.0$), the initial central density, and the critical density of the EOS
($\rho_\mathrm{cr}$).  
The dashed vertical lines represent
the densities at which grid refinement was carried out according to the Jeans condition.
\label{surface_volume_allmodels_f.eps}
}
\end{figure}

Figure~\ref{surface_volume_allmodels_f.eps} compares the surface-to-volume ratio and axis ratios for the three models. Even for the same
initial Mach number, low-density regions with
$\rho \simeq 10^{-20}\,\mathrm{g}\,\mathrm{cm}^{-3}$ 
are more disturbed in the more
massive model, showing a higher surface-to-volume ratio
(Fig.~\ref{surface_volume_allmodels_f.eps}{\it a}).
The less massive model produces a disk-shaped envelope in the range
$10^{-17}\,\mathrm{g}\,\mathrm{cm}^{-3}\lesssim\rho\lesssim10^{-12}\,\mathrm{g}\,\mathrm{cm}^{-3}$, 
indicated by the identical
values of the two axis ratios (red curves in Fig.~\ref{surface_volume_allmodels_f.eps}{\it b}).
In contrast, the massive clouds produce a filamentary
envelope.  The axis ratio $a_3/a_1$ reaches a maximum value of $\simeq 80$ for the most
massive model (blue curves in
Fig.~\ref{surface_volume_allmodels_f.eps}{\it b}).
For $\rho \gtrsim 10^{-10}\,\mathrm{g}\,\mathrm{cm}^{-3}$, 
the surface-to-volume ratio and the axis ratios are almost unity,
irrespective of the cloud masses because of the spherical first cores.

\begin{figure}
\epsscale{1.0}
\plotone{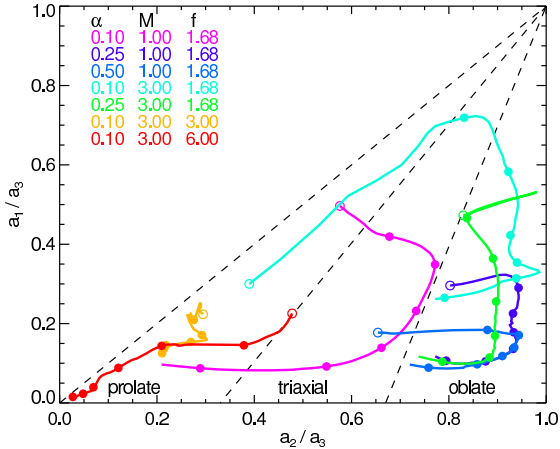}
\figcaption[aratio_seq.eps]{ 
Distribution of axis ratio for the isothermal range ($\rho_0 \le
\rho \le \rho_\mathrm{cr}$) at the stage with
$\rho_\mathrm{max} = 10^{-8} \,\mathrm{g}\,\mathrm{cm}^{-3}$ for all the collapse models with
turbulence (${\cal M} \neq 0$).
The open circles denote the axis ratios at $\rho = \rho_0$.
The filled circles denote the axis ratios at $\rho = 10^n
\rho_0$ ($n = 1, 2, \cdots, 6$).
The dashed lines indicate the boundary between prolate, triaxial, and
oblate shapes.
\label{aratio_seq.eps}
}
\end{figure}

Figure~\ref{aratio_seq.eps} shows the distribution of the axis ratios
of the isothermal envelopes ($\rho_0 \le \rho \le
\rho_\mathrm{cr}$).  Based on the axis ratios, the shapes of
the envelopes are divided into three categories: prolate, triaxial, and
oblate \citep{Gammie03}.  
In all the collapse models, the shape anisotropy increases ($a_1/a_3$ decreases)
with the density threshold for $\rho
\gtrsim 10 \rho_0$ ($=10^{-18}\,\mathrm{g}\,\mathrm{cm}^{-3}$).
All the massive cloud cores with $f = 3.0$ and 6.0 have prolate shapes, with $a_2/a_3$
and $a_1/a_3$ decreasing with increasing density threshold.
Four of the five less massive cloud cores with $f=1.68$
produce oblate cores for $\rho > 100 \rho_0$.
In addition, the evolutionary tracks of the axis ratios for the dense regions ($\rho \ge
\rho_\mathrm{max}/10$) exhibit loci similar to those shown in
Figure~\ref{aratio_seq.eps}, indicating that the shapes of the
envelopes reflect the history of the collapse. In summary, massive cloud cores tend to have a prolate shape, and 
less massive cores an oblate shape.

%% Energy ratio

\begin{figure}
\epsscale{1.0}
\plotone{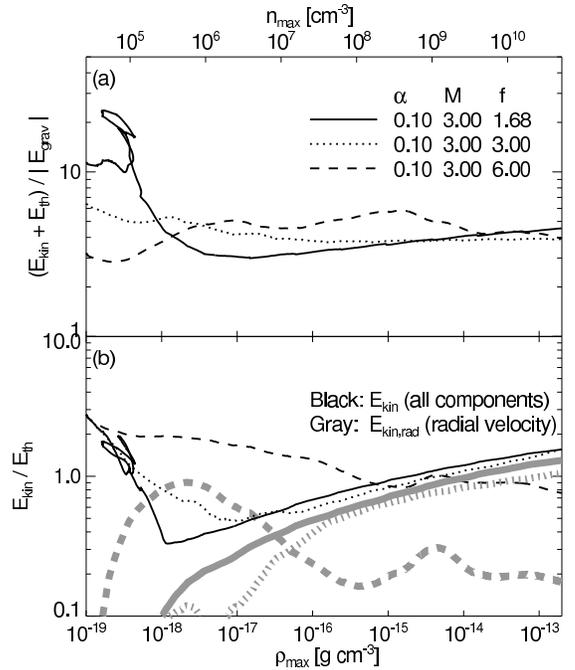}
\figcaption[Eturb_th_rhomax2_bw.eps]{ 
Energy ratios
({\it a}) $(E_\mathrm{kin} + E_\mathrm{th})/|E_\mathrm{grav}|$ and 
({\it b}) $E_\mathrm{kin} / E_\mathrm{th}$ 
 as functions of the maximum density $\rho_\mathrm{max}$
in the isothermal collapse phase
for the models with ${\cal M} =3$ and $\alpha = 0.1$. 
Solid, dotted, and dashed lines correspond to models with
$f = 1.68$, 3.0, and 6.0, respectively.
Gray lines in the panel ({\it b}) denote the ratios of the kinetic energy
of the radial velocity to the thermal energy $E_\mathrm{kin,rad}/E_\mathrm{th}$.
\label{Eturb_th_rhomax2_bw.eps}
}
\end{figure}

In order to investigate the deformation of the cloud cores,
we consider the kinetic, thermal, and gravitational energy  of the dense regions during the collapse.
Figure~\ref{Eturb_th_rhomax2_bw.eps}{\it a} shows 
$(E_\mathrm{kin} + E_\mathrm{th})/|E_\mathrm{grav}|$ in the dense
region with $\rho \ge 0.1 \rho_\mathrm{max}$ during the isothermal
collapse phase ($\rho_0 \le \rho_\mathrm{max} \le \rho_\mathrm{cr}$).
These energies are estimated as 
$E_\mathrm{kin} = (1/2) \int_{\rho \ge 0.1 \rho_\mathrm{max}} \rho |\bmath{v}|^2 dV$, 
$E_\mathrm{th} =  (3/2) \int_{\rho \ge 0.1 \rho_\mathrm{max}} P dV$, 
and
$| E_\mathrm{grav} | =  (1/8\pi G) \int_{\rho \ge 0.1 \rho_\mathrm{max}} |\bmath{g}|^2  dV$.
As shown in Figure~\ref{Eturb_th_rhomax2_bw.eps}{\it a}, 
the values of the energy ratio $(E_\mathrm{kin} +
E_\mathrm{th})/|E_\mathrm{grav}|$ converge to $\sim 4$ by the
early stage with $\rho_\mathrm{max} \simeq
10^{-18}\,\mathrm{g}\,\mathrm{cm}^{-3}$, and
remain roughly constant during the isothermal 
collapse.   
This convergence is explained by the virial theorem, which predicts that
$(E_\mathrm{kin}+E_\mathrm{th}+E_\mathrm{mag}/2)/|E_\mathrm{grav}|
\simeq 1/2$. The overestimation in
Figure~\ref{Eturb_th_rhomax2_bw.eps}{\it a} is attributed to the
underestimation of the gravitational energy, since we ignored the
contribution of the low density regions. 

Figure~\ref{Eturb_th_rhomax2_bw.eps}{\it b} shows the energy ratio, 
$E_\mathrm{kin}/E_\mathrm{th}$, during the isothermal collapse.
For $\rho_\mathrm{max} \lesssim 10^{-15}\,\mathrm{g}\,\mathrm{cm}^{-3}$, 
the kinetic energy exceeds the thermal energy for the most massive
model with $f = 6$ (thin dashed curve), and vice verse for the less
massive model with $f=1.68$ (thin solid curve).
The massive cloud core is supported
mainly by the turbulence, and it collapses to form a filament. 
In contrast, the oblate shape in the less massive model is attributed to
thermal pressure support. 

In the early phase, the energy ratio
$E_\mathrm{kin}/E_\mathrm{th}$ decreases for all the models.
However, they begin to increase at $\rho_\mathrm{max} =
10^{-18}\,\mathrm{g}\,\mathrm{cm}^{-3}$ for the model with $f = 1.68$
and at $10^{-17}\,\mathrm{g}\,\mathrm{cm}^{-3}$ for the model with $f = 3$.
This increase is attributed to an increase in the infall energy,
as indicated by the gray curves, which denote the energy ratio
$E_\mathrm{kin,rad}/E_\mathrm{th}$, where $E_\mathrm{kin,rad}$ is the kinetic
energy associated only with the radial velocity, and is defined by 
$E_\mathrm{kin, rad} = (1/2) \int_{\rho \ge 0.1 \rho_\mathrm{max}} \rho v_r^2 dV$.
For $\rho_\mathrm{max} \gtrsim 10^{-16}\,\mathrm{g}\,\mathrm{cm}^{-3}$, 
the energy of infall motion dominates over the turbulent energy
for the models with $f = 1.68$ and 3.0.

\subsection{Outflow, rotation, and magnetic field}
\label{sec:outflow}

As shown in \S\ref{sec:lowmass} and \S\ref{sec:highmass}, the models
examined here exhibit either bipolar or spiral outflow.  Examples of bipolar flows are shown in 
Figures~\ref{Outflows_allmodel.eps}{\it a},
\ref{Outflows_allmodel.eps}{\it b}, and
\ref{Outflows_allmodel.eps}{\it d}, and spiral flows in 
Figures~\ref{Outflows_allmodel.eps}{\it c},
\ref{allmodel_F.eps}{\it f} and \ref{allmodel_F.eps}{\it i}.
The bipolar outflows tend to be associated with disk-shaped envelopes, while 
the spiral flows tend to have filamentary envelopes.

In order to investigate the relationship between the rotation and magnetic
field, we calculate the angular momentum ($\bmath{J}$) and
the mean magnetic field ($\bar{\bmath{B}}$) for $\rho \ge 0.1
\rho_\mathrm{max}$ during the collapse (see Appendix \ref{sec:flux-spin-relation}). 
The angular momentum can be decomposed into two components parallel and perpendicular
to the mean magnetic field, defined by 
\begin{equation}
J_\parallel = |\bmath{J} \cdot \bar{\bmath{B}}|/|\bar{\bmath{B}}|,
\end{equation}
and 
\begin{equation}
J_\perp = |\bmath{J} \times \bar{\bmath{B}}|/|\bar{\bmath{B}}|.
\end{equation}
It should be pointed out that the region of interest
$\rho \ge 0.1 \rho_\mathrm{max}$ changes temporally.  Therefore, changes in $\bmath{J}$ and $\bar{\bmath{B}}$ do not indicate 
temporal changes in the angular momentum and magnetic field of 
the fixed regions, but rather they trace the angular momentum and magnetic field
in the collapsing region.

\begin{figure}
\epsscale{1.0}
\plotone{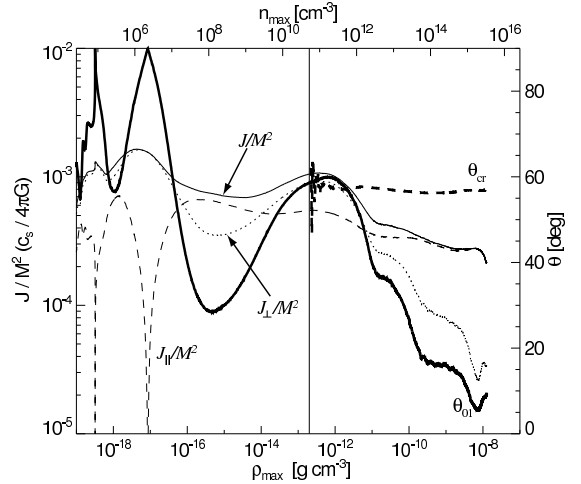}
\figcaption[j_b_angle_M1B025.eps]{ 
Normalized angular momentum ($J/M^2 [c_s/4\pi G]$) within the dense
region with $\rho \ge 0.1 \rho_\mathrm{max}$ as a function of
the maximum density $(\rho_\mathrm{max})$ for the model with 
$(\alpha, {\cal M}, f) = (0.25, 1.0, 1.68)$. 
The angular momentum $J/M^2$ is plotted as a thin solid curve, while
the parallel and perpendicular components of angular momentum
with respective to the local
magnetic field, $J_\parallel/M^2$ and $J_\perp/M^2$, are plotted as 
thin dashed and thin dotted curves, respectively.  
The thick solid curve denotes the angle $\theta_\mathrm{01}$ 
between the vectors of the angular momentum $\bmath{J}$ and the mean local magnetic
field $\bar{\bmath{B}}$ for the region with $\rho \ge 0.1 \rho_\mathrm{max}$, while the thick
dashed curve denotes the angle $\theta_\mathrm{cr}$, which is same as $\theta_\mathrm{01}$ 
but for the region
of $\rho \ge \rho_\mathrm{cr}$.  Both angles are restricted to within
$[0, 90]$ degrees. The vertical line denotes the critical density of the EOS, $\rho_\mathrm{cr}$.
\label{j_b_angle_M1B025.eps}
}
\end{figure}

\begin{figure}
\epsscale{1.0}
\plotone{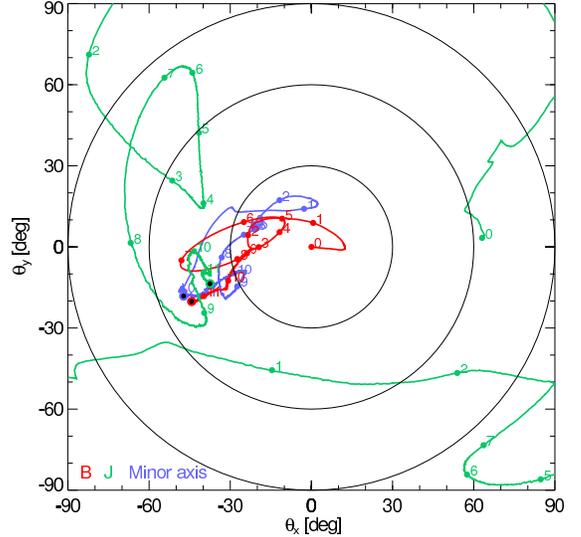}
\figcaption[j_b_it_M1B025.eps]{ 
Loci of the directions of the mean magnetic field $\pm\bar{\bmath{B}}$ (red line), 
the angular momentum $\pm\bmath{J}$ (green line), and the minor axis of 
the density structure (blue line) in the region with $\rho \ge 0.1 \rho_\mathrm{max}$ for the model with 
$(\alpha, {\cal M}, f) = (0.25, 1.0, 1.68)$. 
Digits $n$ with filled circles denote the stages of 
$\rho_\mathrm{max} = 10^n \rho_0$.  
Black filled circles denote the final stage.
The three large circles denote the relationships $(\theta_x^2 + \theta_y^2)^{1/2} =
30^\circ$, $60^\circ$, and $90^\circ$.
\label{j_b_it_M1B025.eps}
}
\end{figure}

Figure~\ref{j_b_angle_M1B025.eps} shows $J/M^2$, $J_\parallel/M^2$, and
$J_\perp/M^2$ as functions of $\rho_\mathrm{max}$ for the model
with $(\alpha, {\cal M}, f) = (0.25, 1.0, 1.68)$, the representative
model of a bipolar outflow.  
The ordinate $J/M^2$ is nearly equal to the specific angular momentum
per unit mass $j/M$, which remains constant as the disk-like
cloud collapses in the cylindrical radial direction 
\citep[see e.g.,][]{Matsumoto97}.
For a slowly rotating non-magnetized cloud, $J/M^2$ increases slightly with density
because of a slight spin-up due to spherical collapse 
 \citep[see Fig.~4 in][]{Matsumoto04}.

In isothermal collapse ($\rho_\mathrm{max} \le \rho_\mathrm{cr}$),
$J/M^2$ decreases slightly with a considerable amount of oscillation.
The slight decrease in the $J/M^2$ is attributed to magnetic
braking.  
In the early phase with $\rho_\mathrm{max} \lesssim 10^{-16}\,\mathrm{g}\,\mathrm{cm}^{-3}$, the
perpendicular component, $J_\perp$, is larger than the parallel
component, $J_\parallel$, and exhibits a large value of $\theta_{01}$, which is the angle between $\bmath{J}$ and
$\bar{\bmath{B}}$, and it is restricted to values from $0$ to $90^\circ$.  The angle
$\theta_{01}$ therefore becomes 0 when
$\bmath{J}$ and $\bar{\bmath{B}}$ are either parallel or anti-parallel.
In the range of $10^{-16}\,\mathrm{g}\,\mathrm{cm}^{-3}
 \lesssim \rho_\mathrm{max} \lesssim
10^{-14}\,\mathrm{g}\,\mathrm{cm}^{-3}$, 
$J_\perp$ is smaller than $J_\parallel$, and
$\theta_{01}$ also becomes smaller.  When the EOS becomes adiabatic ($\rho_\mathrm{max}=
\rho_\mathrm{cr}$), $J_\perp$ is larger than $J_\parallel$.
Also plotted in Figure~\ref{j_b_angle_M1B025.eps} is another angle $\theta_\mathrm{cr}$, which is the angle between the
angular momentum and the magnetic filed for the region with $\rho \ge
\rho_\mathrm{cr}$.
After the EOS becomes adiabatic ($\rho_\mathrm{max} \ge \rho_\mathrm{cr}$), 
the angle $\theta_\mathrm{cr}$ remains constant.
In contrast, $\theta_{01}$ decreases up to $\sim
5^\circ$ by the final stage, indicating that the rotation axis is
aligned with the magnetic field in the dense region.  
Moreover, the angular momentum $J/^2M$ decreases. 
The perpendicular component decreases selectively by more than an order of magnitude.  This
indicates that the magnetic braking is more effective against the
perpendicular component of angular momentum with respective to the
local magnetic field than the parallel component.
Such selective magnetic braking is also reported in \citet{Matsumoto04}.

Figure~\ref{j_b_it_M1B025.eps} shows the loci of the directions of the
mean magnetic field $\bar{\bmath{B}}$, the angular momentum
$\bmath{J}$, and the minor axis of the density structure (the
normal vector of the disk) in the
region with $\rho \ge 0.1 \rho_\mathrm{max}$ for the model with $(\alpha,
{\cal M}, f) = (0.25, 1.0, 1.68)$.  Each vector is plotted in the
two-dimensional plane as 
\begin{equation}
\left(
\begin{array}{c}
\theta_x \\
\theta_y
\end{array}
\right)
 = \frac{\arctan(V_{xy}/V_z)}{V_{xy}}
\left(
\begin{array}{c}
V_x \\
V_y
\end{array}
\right),
\end{equation}
where $\bmath{V} = (V_x, V_y, V_z)$ represents a given vector and
$V_{xy} = (V_x^2+V_y^2)^{1/2}$.  The distance from the origin
represents the angle between a given vector and the $z$-axis. For
example, a vector parallel to the $x$-axis is plotted at $(90^\circ,
0)$.  
The direction of the magnetic field and the minor axis of the density structure
are aligned during the collapse, suggesting that the cloud collapses
along the magnetic field lines, and the disk is oriented always
perpendicular to the local magnetic field.
The direction of the angular
momentum drifts considerably over the $\theta_x - \theta_y$ plane,
while it begins to be aligned with the magnetic field and the minor
axis at $\rho_\mathrm{max}/\rho_0 \sim 10^8$ 
($\rho_\mathrm{max} = 10^{-11}\,\mathrm{g}\,\mathrm{cm}^{-3}$).
Finally, for $\rho_\mathrm{max}/\rho_0 \gtrsim 10^{10}$
($\rho_\mathrm{max} = 10^{-8}\,\mathrm{g}\,\mathrm{cm}^{-3}$), 
all three vectors are aligned at the point 
$(\theta_x, \theta_y) \simeq (-40^\circ, -20^\circ)$.
This alignment produces the bipolar outflow.

\begin{figure}
\epsscale{1.0}
\plotone{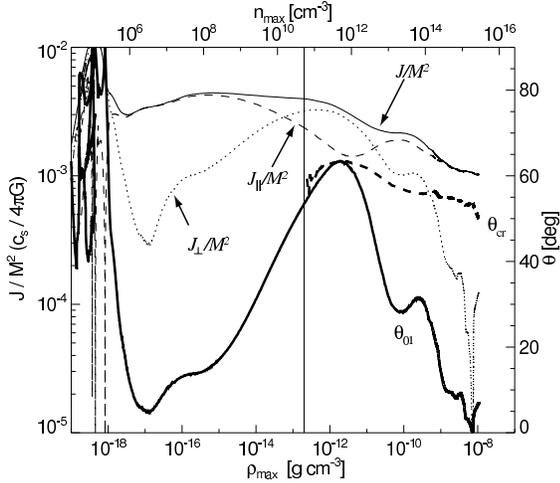}
\figcaption[j_b_angle_M3B01.eps]{ 
Same as figure~\ref{j_b_angle_M1B025.eps} but for 
the model with $(\alpha, {\cal M}, f) = (0.1, 3.0, 1.68)$. 
\label{j_b_angle_M3B01.eps}
}
\end{figure}

\begin{figure}
\epsscale{1.0}
\plotone{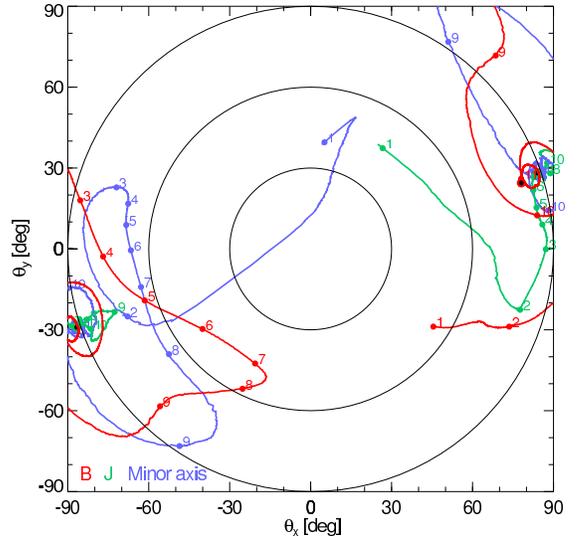}
\figcaption[j_b_it_M3B01.eps]{ 
Same as figure~\ref{j_b_it_M1B025.eps} but for 
the model with $(\alpha, {\cal M}, f) = (0.1, 3.0, 1.68)$. 
The loci are shown for $\rho_\mathrm{max}/\rho_0 \ge
10$, because they are noisy below that stage.
\label{j_b_it_M3B01.eps}
}
\end{figure}

Figure~\ref{j_b_angle_M3B01.eps} is same as
Figure~\ref{j_b_angle_M1B025.eps} but for the model with 
with $(\alpha, {\cal M}, f) = (0.1, 3.0, 1.68)$, where the most
prominent bipolar outflow appears.  
In the early stage with 
$\rho_\mathrm{max} \lesssim 10^{-18}\,\mathrm{g}\,\mathrm{cm}^{-3}$, the angular
momentum strongly oscillates due to the strong
turbulence and the weak magnetic field.  After the collapse
proceeds, $J/M^2$ remains roughly constant in 
the isothermal collapse phase ($\rho_\mathrm{max} \leq \rho_\mathrm{cr}$).
In this phase, the direction of the angular momentum changes considerably as shown
in Figure~\ref{j_b_it_M3B01.eps}.  
The weak magnetic field changes its direction significantly due to precession.
The disk normal follows
the magnetic field, so that the plane of the disk remains perpendicular to the
magnetic field. %
In the adiabatic phase ($\rho_\mathrm{max} \ge \rho_\mathrm{cr}$), 
$J/M^2$ deceases with increasing density as shown in Figure~\ref{j_b_angle_M3B01.eps}. 
The perpendicular component $J_\perp/M^2$ decreases drastically
as in the previous model.
This model therefore results in
alignment of the rotation axis with the local magnetic field, with
$\theta_{01} \sim 0$ in the final stage. 
This alignment is also shown in Figure~\ref{j_b_it_M3B01.eps}, where
the directions of the magnetic field, the angular momentum, and the
disk normal converge at the point 
$(\theta_x, \theta_y) \simeq (80^\circ, 30^\circ)$, 
that is almost perpendicular to the $z$ axis.
The angular momentum $J/M^2$ is about five times larger than that for
the previous model throughout its evolution (compare
Figs.\ref{j_b_angle_M1B025.eps} and \ref{j_b_angle_M3B01.eps}), and this model exhibits strong outflow.  The large angular momentum is attributed to both the
initial strong turbulence and the weak magnetic field.

\begin{figure}
\epsscale{1.0}
\plotone{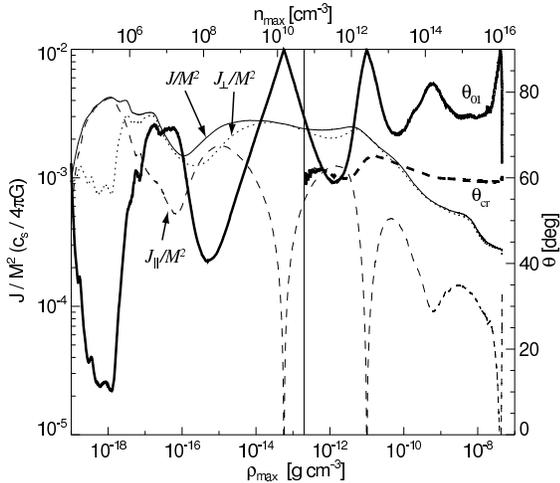}
\figcaption[j_b_angle_M3B01F3.eps]{ 
Same as figure~\ref{j_b_angle_M1B025.eps} but for 
the model with $(\alpha, {\cal M}, f) = (0.1, 3.0, 3.0)$. 
\label{j_b_angle_M3B01F3.eps}
}
\end{figure}

\begin{figure}
\epsscale{1.0}
\plotone{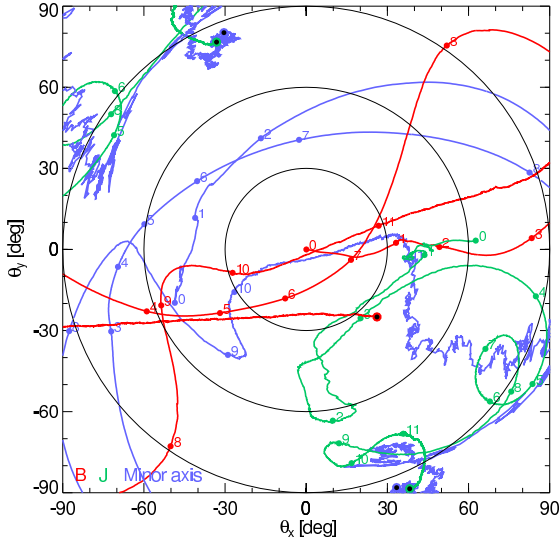}
\figcaption[j_b_it_M3B01F3.eps]{ 
Same as figure~\ref{j_b_it_M1B025.eps} but for 
the model with $(\alpha, {\cal M}, f) = (0.1, 3.0, 3.0)$. 
\label{j_b_it_M3B01F3.eps}
}
\end{figure}

Figure~\ref{j_b_angle_M3B01F3.eps} shows the evolution of angular
momentum for the model with $(\alpha, {\cal M}, f) = (0.1, 3.0,
3.0)$, a typical model showing a spiral flow.
In this model, the angular momentum and magnetic field are not aligned
throughout the evolution.  The angle $\theta_{01}$ increases
and exhibits large oscillations.  The angular momentum $J/M^2$ remains roughly
constant in the isothermal collapse phase ($\rho_\mathrm{max}
\le \rho_\mathrm{cr}$), while it decreases 
in the adiabatic phase ($\rho_\mathrm{max} \ge \rho_\mathrm{cr}$).
The perpendicular component $J_\perp$ is considerably larger than the
parallel component $J_\parallel$ in the adiabatic phase, producing large $\theta_{01}$ and $\theta_\mathrm{cr}$ values.

The magnetic field, the angular momentum, and the minor axis evolve
completely differently in the spiral models as shown in Figure~\ref{j_b_it_M3B01F3.eps}.
The locus of $\bmath{J}$ moves in the lower right direction with a large precession.
The locus of $\bar{\bmath{B}}$ moves diagonally from lower left to upper right, indicating that the orientation of $\bar{\bmath{B}}$ rotates
around $\bmath{J}$.  The point $\bar{\bmath{B}}$ traverses the
$\theta_x-\theta_y$ plane three times, indicating that it
makes 1.5 revolutions.
The minor axis follows the direction of the magnetic field in the stage with
$\rho_\mathrm{max} \lesssim 10^{11}\rho_0$
($\rho_\mathrm{max} \lesssim 10^{-8}\,\mathrm{g}\,\mathrm{cm}^{-3}$), while it follows the
direction of the angular momentum in the stage with
$\rho_\mathrm{max} \gtrsim 10^{11}\rho_0$
($\rho_\mathrm{max} \gtrsim 10^{-8}\,\mathrm{g}\,\mathrm{cm}^{-3}$).
In the final stage, the angular momentum and the minor axis are
aligned at the point $(\theta_x, \theta_y) \simeq (-30^\circ, 80^\circ)$.
This indicates that the filaments shown in
Figure~\ref{allmodel_F.eps}{\it e} and Figure~\ref{wholeF.eps}{\it e}
are perpendicular to the magnetic fields, while the first core
shown in Figure~\ref{wholeF.eps}{\it e} is slightly oblate, with its
minor axis oriented parallel to the rotation axis.

\section{Discussion}
\label{sec:discussion}
\subsection{Shape of cloud core}

\revise{
Cloud core shape is investigated quantitatively by means of the
surface-to-volume ratio and axis ratios
(Figs.~\ref{surface_volume_M1B025.eps},
\ref{surface_volume_allmodels.eps}, and
\ref{surface_volume_allmodels_f.eps}).  In all the models with
turbulence, a cloud core has a complex density distribution before it
begins to collapse.  In contrast, the density distribution is
relatively smooth in a collapsing cloud core, which is partly due to
decay of turbulence before collapse.  Strong turbulence prohibits
cores from collapsing.  In other words, the decay of turbulence
promotes collapse in particular in a less massive core (Fig.~\ref{vnt_time.eps}). 
A dense core begins to collapse when the turbulence decays to a certain
level depending on the core mass.  A less massive core does not
collapse until the turbulence decays to a very low level.  Accordingly
the collapsing cores tend to have a smooth density distribution.  We
also note that it takes a longer time before collapse if the initial
turbulence is stronger.  This is one of the reasons why the density
distribution is relatively smooth in the first cores formed in our
simulations.
}

\revise{
The above argument, however, does not hold for a massive cloud core,
which collapses rapidly before decay of turbulence (Fig.~\ref{Eturb_th_rhomax2_bw.eps}). 
Even when a core forms before the decay of turbulence, it has 
a relatively smooth density distribution.  We suppose that this
is due to shrink of the core.  The diameter of a collapsing core 
is about the Jeans length, which decreases in proportion to 
$ \rho ^{-1/2} $ during the runaway collapse if the gas is
isothermal.  Turbulence supports a core against collapse only
when the typical wavelength is shorter than the diameter of 
the cloud.  Turbulence has a smaller power at a shorter wavelength
in our initial model as well as in the Kolmogorov spectrum.
Accordingly the effective level of turbulence decreases along
with the shrink of the core, unless the wavelength of the 
turbulence shortens in proportion to the Jeans length.
The wavelength of turbulence may be shortened a little 
by compression and advection, but not much.  Remember that
the core mass decreases during the runaway collapse.  This 
means that the advection is slower than the collapse.
The above mentioned idea is supported by the fact that 
the turbulence is still strong in the low density region
even after the collapse (Fig.~\ref{surface_volume_allmodels_f.eps}).
}

Consequently, the turbulence introduces only smooth motion, 
e.g., rotation and shear, into the collapsing region. Our 
results are in agreement with observations of starless dense 
cores with molecular line emissions 
\citep[e.g.,][]{Goodman98,Tafalla02}, which suggest that turbulent motion 
decreases toward the center of the cloud core and coherent 
motion remains at the central part. For star-forming cores, 
the internal structure may be produced by the activation of 
young stars, e.g., outflows from young stars and the orbital 
motion of a multiple system \citep[c.f.,][]{Wang10}.

The shape of the collapsing region depends mainly on the mass of the
cloud core, and the orientation is controlled by the magnetic fields.
In the less massive cloud core, the collapsing region assumes an
oblate shape, whereas in the massive cloud cores, it assumes a prolate
shape. In all cases, the minor axis is parallel to the local magnetic
field and the shape anisotropy increases with density in the
isothermal envelope.

The deformation during the collapse can be explained by the energy
ratio as shown in Figure~\ref{Eturb_th_rhomax2_bw.eps}.  This
analysis indicates that the pressure exerts an isotropic effect, and it
impedes deformation of filaments in the less massive cloud cores.  In
other words, the massive cloud core is subject to few isotropic effects, and it
becomes deformed into a filament.  This is consistent with the results of a classical
analysis by \citet{Lin65}, who investigated deformation during dust cloud collapse. 

There have been many simulation-based studies on large-scale turbulence, in which the
shapes of clumps and cloud cores are have been considered.
\citet{Li04} performed $512^3$ simulations with magnetic
fields, and reported that the majority of cores are prolate or triaxial
in shape.  \citet{Offner09} and \citet{Offner08} performed large-scale AMR
simulations without magnetic fields, and showed that the cores are
predominantly triaxial.  
Our simulations demonstrate that the shape of the cloud core evolves
with increasing density (Fig.~\ref{aratio_seq.eps}), though a
massive core tends to be prolate and a less massive core tends to be oblate.
\citet{Offner09} also suggested that the final shape is somewhat
sensitive to a cutoff density.

We found that the minor axis tends to be aligned with the local
magnetic field irrespective of the shape of the cloud cores (oblate or prolate) in the isothermal collapse phase.  A similar tendency was also
found by \citet{Li04}, who reported 
a weak correlation between the minor axis of the cloud core and the local magnetic field.
Our simulations indicate that the minor
axis is aligned rapidly with the local magnetic field during the
isothermal collapse phase (e.g., Fig.~\ref{j_b_it_M1B025.eps}). 
Moreover, in the weak magnetic field models, 
the minor axis changes orientation during the collapse, 
following the direction of
the magnetic field
(see Figs.~\ref{j_b_it_M3B01.eps}, and \ref{j_b_it_M3B01F3.eps}).
Therefore, the minor axis is always oriented parallel to the local magnetic field.

Hourglass-shaped magnetic fields are reproduced in the present
simulations, as shown in Figures~\ref{avs2000au.eps} and
\ref{avs200au.eps}, which are on 4000~AU and 400~AU scales, respectively.  The
hourglass shapes are prominent in the models with magnetic
fields stronger than or equal to the fiducial strength
($B \gtrsim 20\,\mu\mathrm{G}$ at $n \sim 10^4\,\mathrm{cm}^{-3}$). 
Observations of polarization have revealed such hourglass-shaped
magnetic fields in both high-mass \citep{Schleuning98} and 
low-mass \citep{Sugitani10} star-forming cores. The spatial
scales involved are $\sim 1\,\mathrm{pc}$, which are larger than the scale of the present study.
A notable example is \citet{Girart06}, who resolved the magnetic field
in the low-mass core, NGC 1333 IRAS 4A, on a scale of a few hundred AU.
They revealed hourglass-shaped magnetic fields
perpendicular to the elongated envelope.  However, NGC 1333 IRAS 4A is a binary protostar system.  The possibility of binary formation is discussed in \S\ref{sec:fragmentation}.

\subsection{Outflows}

\begin{figure}
\epsscale{1.0}
\plotone{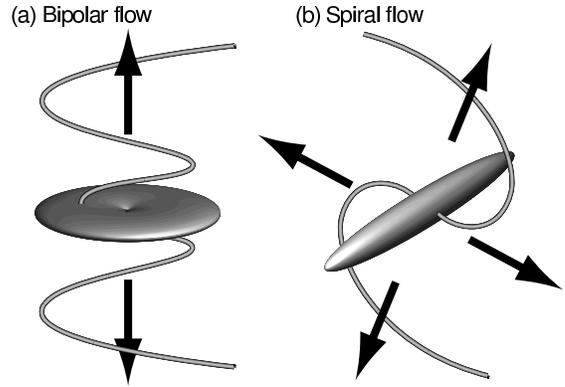}
\figcaption[flows.eps]{ 
Schematic diagram of two types of outflows:
({\it a}) bipolar flow and ({\it b}) spiral flow.
The surfaces represent iso-density surfaces, the tubes denote the
magnetic field lines.  The arrows indicate the direction of the outflows.
\label{flows.eps}
}
\end{figure}

The models here exhibit either bipolar or spiral outflows, as illustrated in Figure~\ref{flows.eps}.  
The former has been produced in both aligned rotators
\citep[i.e.,][]{Tomisaka98,Machida04,Banerjee06,Fromang06} and inclined
rotators \citep[i.e.,][]{Matsumoto04,Hennebelle09}.
The results of our simulations demonstrate that bipolar flows are also produced  in
turbulent cloud cores.

The bipolar flows are divided into two subtypes according to the
configuration of the associated magnetic field.  The outflows in
the moderate field models have a poloidal magnetic field larger than
the toroidal field, as shown in Figures~\ref{Outflows_allmodel.eps}{\it b}
and \ref{Outflows_allmodel.eps}{\it d}.
Another subtype is reported in Figure~\ref{Outflows_allmodel.eps}{\it
  a}, where the toroidal magnetic field dominates over  the
poloidal field.  The former outflow is driven by the
magneto-centrifugal mechanism \citep[c.f.,][]{Blandford82,Pudritz86},
and the latter by the magnetic pressure gradient
of the toroidal field component, categorized as I-type flow in \citet{Tomisaka02}.

The spiral flow is a previously unreported type of outflow. 
The magnetic field lines are wound up by the rotation of the axes oriented perpendicular to the field.  A similar magnetic field morphology was reported by \citet{Machida06} for
inclined rotators (see their Fig.~13).  However, they did
not confirm this type of outflow in their simulations because they
could not follow the evolution for a sufficiently long period.

It is not yet known whether such spiral flows are stable over periods of more
than several thousand years.  When the angular momentum perpendicular to the
magnetic field is released by the outflow, the outflow may transform from
spiral to bipolar.  In order to confirm this possibility, we need to
employ the sink particle method to simulate longer timescales at
a reasonable computational cost.

\subsection{Probability of fragmentation}
\label{sec:fragmentation}

\begin{figure}
\epsscale{1.0}
\plotone{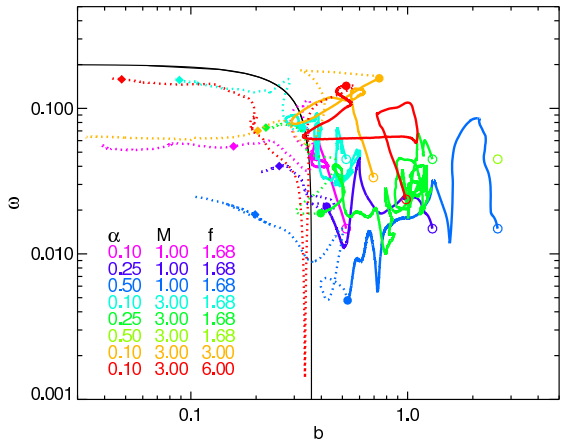}
\figcaption[outflow_asp_grid_only.eps]{ 
Loci in the magnetic flux-spin relation plane for the models with ${\cal M}>0$.
The abscissa and ordinate denote non-dimensional parameters of 
magnetic flux ($b$) and spin ($\omega$), respectively.
Open circles, filled circles, and  diamonds indicate
the stages of $\rho = \rho_0$ (initial stage), 
$\rho_\mathrm{cr}$, and $\rho = 10^{11}\rho_0$.
Loci with continuous curves are for the
isothermal phase ($\rho_\mathrm{max} \leq \rho_\mathrm{cr}$), and loci with
dotted curves are for the adiabatic phase ($\rho_\mathrm{max} >
\rho_\mathrm{cr}$).  
The quarter oval curve indicates the convergence curve, given by
$(b/0.36)^2+(\omega/0.2)^2 =1$.
\label{machidaDiagramPlot.eps}
}
\end{figure}

Figure~\ref{machidaDiagramPlot.eps} displays the evolution of magnetic flux-spin relations for the models with 
${\cal M}>0$.  The magnetic flux-spin relations were proposed by 
\citet{Machida05a,Machida05b} as a means of investigating fragmentation
during cloud collapse.
The abscissa and ordinate denote non-dimensional parameters of 
magnetic flux $b = |\bar{\bmath{B}}| / (8 \pi c_p^2 \bar{\rho})^{1/2}$ and 
spin $\omega = |\bar{\bmath{\Omega}}| / (4 \pi G \bar{\rho})^{1/2}$,
respectively,  for the dense region
with $\rho \ge 0.1 \rho_\mathrm{max} $ (see Appendix~\ref{sec:flux-spin-relation}).  This diagram predicts
fragmentation driven by rotation.

The solid curves in Figure~\ref{machidaDiagramPlot.eps} are loci in the
isothermal phase ($\rho_\mathrm{max} < \rho_\mathrm{cr}$) for all the
turbulent models.  Starting from the
initial condition (open circles), all the loci drift in the $b-\omega$
plane, until they reach the convergence curve denoted by the black curve.
This convergence was also seen for the case of collapse in the absence of turbulence
\citep{Machida05a,Machida05b}.  
They summarized the conditions for fragmentation as follows.  If the
locus reaches the horizontal part of the convergence curve (rotation
dominant) during the isothermal collapse, the cloud core fragments.
If it reaches the vertical part (magnetic field dominant), it does not.
All the models examined here reach the vertical part of the
convergence curve in the isothermal collapse phase, and they do not undergo fragmentation.  
This indicates that turbulent cloud cores have insufficient angular momentum to fragment. 

Another possibility for fragmentation is turbulent fragmentation.  
In the massive models, the turbulence exceeds the thermal pressure as a
supporting force against gravity.  These models produce thin filaments.  
Each filament produces only one first core in the present
simulations, because of the very small time steps  in the adiabatic phase.
However, the formed filaments are very thin and tend to
undergo additional fragmentation in the later stages.  Such
fragmentation could be reproduced if sink particles were introduced into
the simulations. 

After the maximum density exceeds the critical density
$\rho_\mathrm{cr}$, the loci traverse the convergence curve and they
move in the leftward direction (dotted curves).
This leftward movement is attributed to an increase in
the sound speed $c_p$ in the adiabatic phase ($\rho_\mathrm{max} > \rho_\mathrm{cr}$).
Some loci move close to the horizontal part of the convergence curve,
where rotation support exceeds magnetic field support.  Based on
these models, a protoplanetary disk may fragment in the later stages
if it is sufficiently massive compared to its central star.

\section{Summary}
\label{sec:summary}

The collapse of turbulent magnetized cloud cores is investigated by
AMR simulations, resolving both the cloud core and the first core.

The cloud core has a complex density distribution in the low density region corresponding to its
boundary with the parent cloud, because of disturbances due to turbulence.
After the collapse begins, the density distribution becomes smooth in
the collapsing region. 
This indicates that the collapse dilutes
the tiny fluctuations caused by the turbulence.
Even after formation of the first core, the edge of the cloud core remains turbulent.

The shape anisotropy of the collapsing region increases 
during isothermal collapse and depends mainly on the mass.
When a cloud core is less massive ($f = 1.68$), the dense region of
the cloud core is oblate, with its minor axis parallel to the local
magnetic field.  The collapsing region is threaded by an hourglass-shaped distribution of magnetic field lines on a scale of $\lesssim 1000$~AU.
When a cloud core is massive ($f = 3.0$ and 6.0),
the dense region of the cloud core is prolate, with
the minor axis parallel to the local magnetic field.  
The extremely massive cloud ($f=6.0$) produces a very thin filament, which may
later fragment.

In all cases, the orientation of the cloud core is controlled by
the magnetic field.  The minor axis of the collapsing region changes
direction during the collapse, following the direction of the
local magnetic field irrespective of the cloud core shape.
Each model produces a spherical first core, in which 
the density is higher than $\simeq 10^{-10} \mathrm{g\,cm}^{-3}$.
The first core is embedded in the infalling envelope.

The dependence of the shape on the mass can explained by the energy ratio
between the thermal pressure and the turbulence. When the energy due to
thermal pressure exceeds that due to turbulence, the cloud core
becomes oblate in the early stages of collapse.  If the opposite is the case,
then the cloud core assumes a filamentary shape.

We found two types of outflows: bipolar and spiral flows.
Bipolar flow is associated with the disk-shaped envelope, while the
spiral flow is associated with the filamentary envelope.
Bipolar flow tends to occur in less massive cloud cores, and the rotation axis, magnetic field, and the disk normal become aligned in the proximity of the first core.
The disk-outflow system is inclined completely
from the global magnetic field for the weak
magnetized cloud core ($B = 7 \mu \mathrm{G}$). Even for the moderate field models ($B = 19 \mu \mathrm{G}$), 
the outflow direction diverges considerably from that of the global magnetic field 
by an angle of $\sim 40-50^\circ$. For the strong field model ($B = 37 \mu \mathrm{G}$), the disk
envelope is aligned with its minor axis along the direction of the global magnetic field, while the outflow is not produced. The angular momentum perpendicular to
the local magnetic field is selectively reduced in the dense region by
magnetic braking and the outflow during the adiabatic phase in the
bipolar flow models.  
The spiral flow tends to be associated with massive cloud cores, and the rotation axis is not aligned with the magnetic field in the envelope surrounding the first core.  

\acknowledgments 

Numerical computations were carried out on a Cray XT4 at the Center for
Computational Astrophysics (CfCA) at the National Astronomical Observatory
of Japan.
This research was supported in part by a
Grant-in-Aid 
for Scientific Research (C) 20540238
and (B) 22340040
from the Ministry of Education, Culture, Sports, Science and Technology, Japan.

%\appendix
\begin{appendix}
\section{Energies of the critical Bonner-Ebert sphere}
\label{seq:energies_of_be}
The gravitational potential $\Phi$ of the Bonner-Ebert sphere is obtained by
solving the equations \citep[see][]{Chandrasekhar39}
\begin{eqnarray}
\frac{dw}{d\xi} &=& y\\
\frac{dy}{d\xi} &=& -\frac{2y}{\xi} + e^{-w}, 
\end{eqnarray}
imposing the boundary conditions of $w(0) = 0$ and $y(0) = 0$,
where the non-dimensional variables $\xi$ and $w$ are defined by
\begin{eqnarray}
\xi  &=& \frac{r}{a}, \\
w &=& \frac{\Phi}{c_s^2} = - \ln \left(\frac{\rho}{\rho_0}\right), \\
a &=& \left( \frac{c_s^2}{4 \pi G \rho_0} \right)^{1/2}.
\end{eqnarray}
The critical Bonner-Ebert sphere has a maximum radius of
$\xi_\mathrm{max} = 6.45$ ($r_\mathrm{max} = 6.45a$).
The central gravitational potential is equal to zero, $\Phi(0) = 0$,
owing to the boundary conditions.
Considering an isolated critical Bonner-Ebert sphere, 
the gravitational potential in $r \le r_\mathrm{max}$ is given by
\begin{equation}
\Psi(r) = \Phi(r)-\Phi(r_\mathrm{max}) - G\frac{M(r_\mathrm{max})}{r_\mathrm{max}},
\end{equation}
where $M(\xi)$ denotes the mass inside the radius $\xi = r/a$,
\begin{equation}
M(\xi) = 4\pi \int_0^r \rho r^2 dr = 4\pi a^3 \rho_0 y(\xi) \xi^2.
\end{equation}
The gravitational potential is therefore obtained as
\begin{equation}
\Psi(r) = c_s^2 \left[ w(\xi) - w(\xi_\mathrm{max}) - 
\xi_\mathrm{max} y(\xi_\mathrm{max}) \right].
\end{equation}
The gravitational energy is given by
\begin{eqnarray}
E_\mathrm{grav} &=& \frac{1}{2}\int_{r \le r_\mathrm{max}} \rho \Psi dV\\
&=& 2 \pi \rho_0 c_s^2 a^3 \left\{ \int_0^{\xi_\mathrm{max}}  w e^{-w} \xi^2 d\xi 
- \left[\xi^2y(w + \xi y)\right]_{\xi=\xi_\mathrm{max}} \right\}
\\
&=&-352 \rho_0 c_s^2 a^3
\end{eqnarray}
The thermal energy is given by
\begin{eqnarray}
E_\mathrm{th} &=& \frac{3}{2}\int_{r \le r_\mathrm{max}} P dV \\
&=& 6\pi a^3 \rho_0 c_s^2 \left( y \xi^2 \right)_{\xi =\xi_\mathrm{max}}\\
&=& 296 \rho_0 c_s^2 a^3 .
\end{eqnarray}
The magnetic field is expressed as
$B = 2 \pi \alpha G^{1/2} \Sigma$ in our models, where the central surface density
is given by
\begin{eqnarray}
\Sigma &=& 2 \int_0^{r_\mathrm{max}} \rho dr\\
&=& 2 \rho_0 a \int_0^{\xi_\mathrm{max}} e^{-w} d \xi\\
&=& 5.38 \rho_0 a.
\end{eqnarray}
The magnetic energy is therefore given by
\begin{eqnarray}
E_\mathrm{mag} &=& \int_{r \le r_\mathrm{max}} \frac{B^2}{8\pi} dV\\
&=& \frac{2\pi}{3} \rho_0 c_s^2 a^3 \alpha^2 \xi_\mathrm{max}^3 \left(
  \int_0^{\xi_\mathrm{max}} e^{-w} d\xi\right)^2\\
&=& 4061 \alpha^2 \rho_0 c_s^2 a^3.
\end{eqnarray}
The kinetic energy is evaluated as
\begin{equation}
E_\mathrm{kin} =  \frac{1}{2}\int_{r \le r_\mathrm{max}} \rho \bmath{v}^2 dV.
\end{equation}
The initial velocity field is generated using a random number.
When ten velocity fields are generated by changing the 
the seed of the random number,
we obtained $E_\mathrm{kin}$ in the range of $(85.1-419)\rho_0 c_s^2 a^3 {\cal M}^2$
and with an average of $220\rho_0 c_s^2 a^3{\cal M}^2$.
For the velocity field used in this paper, the kinetic energy is
evaluated as $E_\mathrm{kin} = 139 \rho_0 c_s^2 a^3{\cal M}^2$.

% ekin = 211.60968  seed1
% ekin = 328.26955  seed2
% ekin = 281.07859  seed3
% ekin = 187.11815  seed4
% ekin = 157.77626  seed5
% ekin = 187.68924  seed6
% ekin = 182.06888  seed7
% ekin = 85.136322  seed8
% ekin = 419.75509  seed9
% ekin = 159.93797  seed10

Introducing the parameters of the density enhancement factor $f$ (see
equation [\ref{eq:density-enhancement}]), we obtain scaling lows of 
$E_\mathrm{th} \propto f^{3/2}$, 
$E_\mathrm{kin} \propto f^{3/2} {\cal M}^2$, 
$E_\mathrm{grav} \propto f^{5/2}$, 
$E_\mathrm{mag} \propto f^{5/2} \alpha^2$, which yield
energy ratios of 
$E_\mathrm{th}/|E_\mathrm{grav}| = 0.836 f^{-1}$,
$E_\mathrm{kin}/|E_\mathrm{grav}| = 0.394 f^{-1}{\cal M}^2$,
and $E_\mathrm{mag}/|E_\mathrm{grav}| = 11.5\alpha^2$.

\section{Surface-to-volume ratio and axis ratios}
\label{sec:surface-to-volume}
In order to estimate the complexity of the density structure,
the normalized surface-to-volume ratio is calculated.  
We calculate the surface and volume of a region $\Omega (\rho)$ 
where the density is larger than a given threshold $\rho$ and the point of
maximum density is included.
Based on Minkowski functional analysis, 
the volume $V(\rho)$ and surface $S(\rho)$ are calculated by
\begin{eqnarray}
V(\rho) &=& N_0 \Delta V,\\
S(\rho) &=& (6N_0 - 2N_1) \Delta S, \\
N_0 &=& \sum_{i,j,k} n_{i,j,k} ,\\
N_1 &=& \sum_{i,j,k} 
\left( 
n_{i,j,k}  n_{i+1,j,k}
+ n_{i,j,k}  n_{i,j+1,k}
+ n_{i,j,k}  n_{i,j,k+1}
\right)  ,\\
n_{i,j,k} &=& \left\{ 
\begin{array}{ll}
1 & \mathrm{for \; a \; cell \;in \;} \Omega(\rho)\\
0 & \mathrm{otherwise}
\end{array}
\right. ,
\end{eqnarray}
where the subscripts $i,j,k$ denote the cell numbers in the $x$, $y$, and
$z$-directions, respectively.  The symbols $\Delta V$ and $\Delta S$
denote the volume and surface of a cell, respectively.

The surface-to-volume ratio $S(\rho)/V(\rho)^{2/3}$
has a large value when the region $\Omega(\rho)$ has a complex shape.
The ratio  $S(\rho)/V(\rho)^{2/3}$ has a minimum
value when the region $\Omega(\rho)$ is spherical.
We normalize the surface-to-volume ratio so that its value is
unity for a sphere.  Because the computational cell
is cubic, the surface area of a sphere is given approximately by $6 \pi r^2$ 
instead of $4 \pi r^2$, where $r$ denotes the radius of the sphere.
The volume of the sphere is approximated by $4\pi r^3/3$.
The normalized surface-to-volume ratio is therefore given by
\begin{equation}
R(\rho) 
= \frac{2^{4/3}}{3^{2/3} 6 \pi^{1/3}}
\frac{S(\rho)}{V(\rho)^{2/3}}
= 0.138 \frac{S(\rho)}{V(\rho)^{2/3}}.
\end{equation}

We also measure the lengths of the principal axes of the region
$\Omega(\rho)$, calculating the moment of the coordinates
\begin{equation}
K_{\hat{i},\hat{j}} = \frac{1}{V(\rho)}\sum_{i,j,k} n_{i,j,k}
(r_{i,j,k,\hat{i}} - r_{g,\hat{i}})(r_{i,j,k,\hat{j}} - r_{g,\hat{j}})
\Delta V,
\end{equation}
where $\bmath{r}_{i,j,k} = (r_{i,j,k,1}, r_{i,j,k,2}, r_{i,j,k,3}) =
(x_{i,j,k}, y_{i,j,k}, z_{i,j,k})$ represents the coordinates of the cell.
The coordinates of the baricenter $\bmath{r}_g$ are estimated by
\begin{equation}
\bmath{r}_g = \frac{1}{V(\rho)} \sum_{i,j,k} n_{i,j,k}
\bmath{r}_{i,j,k} \Delta V.
\end{equation}
The three principal axes $a_n$ are defined by the square roots of
three eigenvalues of $K_{\hat{i},\hat{j}}$,
arranged in ascending order ($a_1 < a_2 < a_3$).
The axis ratios are estimated as $a_2/a_1$ and $a_3/a_1$.

When the region $\Omega$ overlaps the boundary of the computational
domain, care must be taken regarding the periodic boundary conditions.  Moreover, 
the axis ratios are obtained only when the one-dimensional length of
the region $\Omega$ is smaller
than the width of the computational domain.

\section{Analysis of dense region}
\label{sec:flux-spin-relation}

We calculate the mean values of the dense region of the cloud cores, 
$\bar{\rho}$, $\bar{\bmath{B}}$, $\bmath{J}$,
$\bar{\bmath{\Omega}}$, and the direction of the minor axis.
The mean density of the dense region is defined by $\bar{\rho}
= M/V$, where
\begin{equation}
M = \int_{\rho \ge 0.1 \rho_\mathrm{max}} \rho(\bmath{r}) dV,
\end{equation}
and
\begin{equation}
V = \int_{\rho \ge 0.1 \rho_\mathrm{max}} dV.
\end{equation}
The integral $\int_{\rho \ge 0.1 \rho_\mathrm{max}}  dV$ denotes a volume
integral over the region with $\rho \ge 0.1 \rho_\mathrm{max}$. 
The mean magnetic field in the dense region is defined by
\begin{equation}
\bar{\bmath{B}} = \frac{1}{V}\int_{\rho \ge 0.1 \rho_\mathrm{max}} \bmath{B}(\bmath{r}) dV.
\end{equation}
The angular momentum in the dense region is defined by
\begin{equation}
\bmath{J} = 
\int_{\rho \ge 0.1 \rho_\mathrm{max}}
(\bmath{r}-\bmath{r}_g)\times \left[ \bmath{v}(\bmath{r}) - \bmath{v}_g\right]  \rho(\bmath{r}) dV,
\end{equation}
where $\bmath{r}_g$ and $\bmath{v}_g$ denote the 
coordinates and velocity of the baricenter, estimated respectively as
\begin{equation}
\bmath{r}_g = \frac{1}{M}\int_{\rho \ge 0.1 \rho_\mathrm{max}} \bmath{r}
\rho(\bmath{r}) dV,
\end{equation}
and
\begin{equation}
\bmath{v}_g = \frac{1}{M}\int_{\rho \ge 0.1 \rho_\mathrm{max}} \bmath{v}(\bmath{r})
\rho(\bmath{r}) dV.
\end{equation}
The mean angular velocity in the dense region is defined by
\begin{equation}
\bar{\bmath{\Omega}} = \bar{\bmath{I}}^{-1} \bmath{J},
\end{equation}
where $\bar{\bmath{I}}$ indicates the moment of inertia, of which
the components are estimated as
\begin{equation}
\bar{I}_{ij} = \int_{\rho \ge 0.1 \rho_\mathrm{max}} \left[ 
(\bmath{r} - \bmath{r}_g) ^2 \delta_{ij}  - (r_i-r_{g,i}) (r_j-r_{g,j}) 
\right]
\rho(\bmath{r}) dV,
\end{equation}
and the subscripts $i$,
$j$ represent coordinate labels, i.e., $x = r_1$, $y=r_2$,
$z=r_3$, 
and $\delta_{ij}$ denotes the Kronecker delta.

The axes of the dense region are derived from the moment of the coordinates,
\begin{equation}
\bar{K}_{i,j} = \int_{\rho \ge 0.1 \rho_\mathrm{max}} (r_i - r_{g,i})(r_j -
r_{g,j}) \rho(\bmath{r}) dV.
\end{equation}
The eigenvectors of $\bar{K}_{i,j}$ indicate the orientation of the
principal axes.

\revise{
The velocity dispersion is given by
\begin{equation}
\left< \Delta v \right> = 
\left\{
\frac{1}{M} 
\int_{\rho \ge 0.1 \rho_\mathrm{max}} 
\left[\bmath{v}(\bmath{r})-\bmath{v}_g\right]^2 \rho(\bmath{r})
dV
\right\}^{1/2}.
\label{eq:delta_v}
\end{equation}
The velocity dispersion for the radial velocity is also estimated as
\begin{equation}
\left< \Delta v_r \right> = 
\left[
\frac{1}{M} 
\int_{\rho \ge 0.1 \rho_\mathrm{max}} 
v_r(\bmath{r})^2 \rho(\bmath{r})
dV
\right]^{1/2},
\label{eq:delta_vr}
\end{equation}
where $v_r(\bmath{r})$ denotes the radial component of the relative
velocity, $\bmath{v}(\bmath{r})-\bmath{v}_g$,
with respective to the position of the baricenter, $\bmath{r}_g$.
}

\end{appendix}


\begin{thebibliography}{}
\bibitem[Banerjee \& Pudritz(2006)]{Banerjee06} Banerjee, R., \& Pudritz, R.~E.\ 2006, \apj, 641, 949 
\bibitem[Bate(2009)]{Bate09} Bate, M.~R.\ 2009, \mnras, 397, 232 
\bibitem[Blandford \& Payne(1982)]{Blandford82} Blandford, R.~D., \& Payne, D.~G.\ 1982, \mnras, 199, 883 
\bibitem[Bonnor(1956)]{Bonnor1956} Bonnor, W. B. 1956, \mnras, 116,351
\bibitem[Burkert \& Bodenheimer(2000)]{Berkert00} Burkert, A., \& Bodenheimer, P.\ 2000, \apj, 543, 822 
\bibitem[Chandrasekhar(1939)]{Chandrasekhar39} Chandrasekhar, S.\ 1939, Chicago, Ill., The University of Chicago press [1939],  
\bibitem[Commer{\c c}on et al.(2010)]{Commercon10} Commer{\c c}on, B., Hennebelle, P., Audit, E., Chabrier, G., \& Teyssier, R.\ 2010, \aap, 510, L3 
\bibitem[Crutcher(1999)]{Crutcher99} Crutcher, R.~M.\ 1999, \apj, 520, 706 
\bibitem[Dedner et al.(2002)]{Dedner02} Dedner, A., Kemm, F., Kr{\"o}ner, D., Munz, C.-D., Schnitzer, T., \& Wesenberg, M.\ 2002, Journal of Computational Physics, 175, 645 
\bibitem[Dubinski et al.(1995)]{Dubinski95} Dubinski, J., Narayan, R., \& Phillips, T.~G.\ 1995, \apj, 448, 226 
\bibitem[Ebert(1955)]{Ebert1955} Ebert, R. 1955, Z. Astrophys., 37, 222
\bibitem[Fromang et al.(2006)]{Fromang06} Fromang, S., Hennebelle, P., \& Teyssier, R.\ 2006, \aap, 457, 371 
\bibitem[Gammie et al.(2003)]{Gammie03} Gammie, C.~F., Lin, Y.-T., Stone, J.~M., \& Ostriker, E.~C.\ 2003, \apj, 592, 203 
\bibitem[Girart et al.(2006)]{Girart06} Girart, J.~M., Rao, R., \& Marrone, D.~P.\ 2006, Science, 313, 812
\bibitem[Goodwin et al.(2004)]{Goodwin04} Goodwin, S.~P., Whitworth, A.~P., \& Ward-Thompson, D.\ 2004, \aap, 414, 633 
\bibitem[Goodman et al.(1998)]{Goodman98} Goodman, A.~A., Barranco, J.~A., Wilner, D.~J., \& Heyer, M.~H.\ 1998, \apj, 504, 223 
\bibitem[Hennebelle \& Ciardi(2009)]{Hennebelle09} Hennebelle, P., \& Ciardi, A.\ 2009, \aap, 506, L29 
\bibitem[Henning et al.(2001)]{Henning01} Henning, T., Wolf, S., Launhardt, R., \& Waters, R.\ 2001, \apj, 561, 871 
\bibitem[Jijina et al.(1999)]{Jijina99} Jijina, J., Myers, P.~C., \& Adams, F.~C.\ 1999, \apjs, 125, 161 
\bibitem[Klein et al.(2007)]{Klein07} Klein, R.~I., Inutsuka, S.-I., Padoan, P., \& Tomisaka, K.\ 2007, Protostars and Planets V, 99 
\bibitem[Langer et al.(1995)]{Langer95} Langer, W.~D., Velusamy, T., Kuiper, T.~B.~H., Levin, S., Olsen, E., \& Migenes, V.\ 1995, \apj, 453, 293 
\bibitem[Larson(1969)]{Larson69} Larson, R.\ B.\ 1969, \mnras, 145, 271 
\bibitem[Larson(1981)]{Larson81} Larson, R.~B.\ 1981, \mnras, 194, 809 
\bibitem[Li et al.(2004)]{Li04} Li, P.~S., Norman, M.~L., Mac Low, M.-M., \& Heitsch, F.\ 2004, \apj, 605, 800 
\bibitem[Lin et al.(1965)]{Lin65} Lin, C.~C., Mestel, L., \& Shu, F.~H.\ 1965, \apj, 142, 1431 
\bibitem[Machida et al.(2004)]{Machida04} Machida, M.~N., Tomisaka, K., \& Matsumoto, T.\ 2004, \mnras, 348, L1 
\bibitem[Machida et al.(2005a)]{Machida05a} Machida, M.~N., Matsumoto, T., Tomisaka, K., \& Hanawa, T.\ 2005, \mnras, 362, 369 
\bibitem[Machida et al.(2005b)]{Machida05b} Machida, M.~N.,Matsumoto, T., Hanawa, T., \& Tomisaka, K.\ 2005, \mnras, 362, 382 
\bibitem[Machida et al.(2006)]{Machida06} Machida, M.~N., Matsumoto, T., Hanawa, T., \& Tomisaka, K.\ 2006, \apj, 645, 1227 
\bibitem[Masunaga, Miyama, \& Inutsuka(1998)]{Masunaga98} Masunaga, H., Miyama, S.~M., \& Inutsuka, S.\ 1998, \apj, 495, 346. 
\bibitem[Matsumoto \& Tomisaka(2004)]{Matsumoto04} Matsumoto, T., \& Tomisaka, K.\ 2004, \apj, 616, 266 
\bibitem[Matsumoto et al.(1997)]{Matsumoto97} Matsumoto, T., Hanawa, T., \& Nakamura, F.\ 1997, \apj, 478, 569 
\bibitem[Matsumoto(2007)]{Matsumoto07} Matsumoto, T.\ 2007, \pasj, 59, 905 
\bibitem[McKee \& Ostriker(2007)]{McKee07} McKee, C.~F., \& Ostriker, E.~C.\ 2007, \araa, 45, 565 
\bibitem[Mellon \& Li(2008)]{Mellon08} Mellon, R.~R., \& Li, Z.-Y.\ 2008, \apj, 681, 1356 
\bibitem[Mellon \& Li(2009)]{Mellon09} Mellon, R.~R., \& Li, Z.-Y.\ 2009, \apj, 698, 922 
\bibitem[Myers et al.(1991)]{Myers91} Myers, P.~C., Fuller, G.~A., Goodman, A.~A., \& Benson, P.~J.\ 1991, \apj, 376, 561 
\bibitem[Nakamura \& Li(2008)]{Nakamura08} Nakamura, F., \& Li, Z.-Y.\ 2008, \apj, 687, 354 
\bibitem[Nakano \& Nakamura(1978)]{Nakano78} Nakano, T., \& Nakamura, T.\ 1978, \pasj, 30, 671 
\bibitem[Offner \& Krumholz(2009)]{Offner09} Offner, S.~S.~R., \& Krumholz, M.~R.\ 2009, \apj, 693, 914 
\bibitem[Offner et al.(2008)]{Offner08} Offner, S.~S.~R., Klein, R.~I., \& McKee, C.~F.\ 2008, \apj, 686, 1174 
\bibitem[Price \& Bate(2008)]{Price08} Price, D.~J., \& Bate, M.~R.\ 2008, \mnras, 385, 1820 
\bibitem[Pudritz \& Norman(1986)]{Pudritz86} Pudritz, R.~E., \& Norman, C.~A.\ 1986, \apj, 301, 571 
\bibitem[Ryden(1996)]{Ryden96} Ryden, B.~S.\ 1996, \apj, 471, 822 
\bibitem[Shu \& Li(1997)]{Shu97} Shu, F.~H., \& Li, Z.-Y.\ 1997, \apj, 475, 251 
\bibitem[Schleuning(1998)]{Schleuning98} Schleuning, D.~A.\ 1998, \apj, 493, 811 
\bibitem[Sugitani et al.(2010)]{Sugitani10} Sugitani, K., et al.\ 2010, \apj, 716, 299 
\bibitem[Tafalla et al.(2002)]{Tafalla02} Tafalla, M., Myers, P.~C., Caselli, P., Walmsley, C.~M., \& Comito, C.\ 2002, \apj, 569, 815 
\bibitem[Tomida et al.(2010)]{Tomida10} Tomida, K., Tomisaka, K., Matsumoto, T., Ohsuga, K., Machida, M.~N., \& Saigo, K.\ 2010, \apjl, 714, L58 
\bibitem[Tomisaka et al.(1988)]{Tomisaka88} Tomisaka, K., Ikeuchi, S., \& Nakamura, T.\ 1988, \apj, 335, 239 
\bibitem[Tomisaka(1998)]{Tomisaka98} Tomisaka, K.\ 1998, \apjl, 502, L163 
\bibitem[Tomisaka(2002)]{Tomisaka02} Tomisaka, K.\ 2002, \apj, 575, 306 
\bibitem[Truelove et al.(1997)]{Truelove97} Truelove, J.~K., Klein, R.~I., McKee, C.~F., Holliman, J.~H., II, Howell, L.~H., \& Greenough, J.~A.\ 1997, \apjl, 489, L179 
\bibitem[Vall{\'e}e et al.(2003)]{Valle03} Vall{\'e}e, J.~P., Greaves, J.~S., \& Fiege, J.~D.\ 2003, \apj, 588, 910 
\bibitem[Wang et al.(2010)]{Wang10} Wang, P., Li, Z.-Y., Abel, T., \& Nakamura, F.\ 2010, \apj, 709, 27 
\bibitem[Wolf et al.(2003)]{Wolf03} Wolf, S., Launhardt, R., \& Henning, T.\ 2003, \apj, 592, 233 
\bibitem[Zuckerman \& Evans(1974)]{Zuckerman74} Zuckerman, B., \& Evans, N.~J., II 1974, \apjl, 192, L149 
\end{thebibliography}
\end{document}